\documentclass[manuscript]{acmart}
\AtBeginDocument{%
  }

\setcopyright{acmlicensed}
\copyrightyear{2018}
\acmYear{2018}
\acmDOI{XXXXXXX.XXXXXXX}
\acmConference[Conference acronym 'XX]{Make sure to enter the correct
  conference title from your rights confirmation email}{June 03--05,
  2018}{Woodstock, NY}
\acmISBN{978-1-4503-XXXX-X/2018/06}

\usepackage{natbib}
\usepackage{csquotes}
\usepackage{hhline}
\usepackage{nicematrix} 
\usepackage{booktabs}   
\usepackage{multirow}   
\usepackage{array}
\usepackage{enumitem}
\usepackage{tabularx}
\usepackage{tablefootnote}
\usepackage{color}
\usepackage{makecell}
\usepackage{colortbl}
\definecolor{tablegray}{gray}{0.95}

\usepackage[table]{xcolor}
\usepackage[normalem]{ulem}
\usepackage{etoolbox}
\usepackage{amsmath}
\usepackage{threeparttable}

\newboolean{showchanges}
\setboolean{showchanges}{false} 
\DeclareRobustCommand{\revAdd}[1]{%
    \ifbool{showchanges}{\textcolor{blue}{#1}}{#1}%
}
\DeclareRobustCommand{\revDel}[1]{%
    \ifbool{showchanges}{\textcolor{red}{\protect\sout{#1}}}{}%
}

\begin{document}

\title[Unpacking the Impact of Real-Time AI Interviewing Assistance]{The Interviewer’s Perspective: Unpacking the Impact of Real-Time AI Interviewing Assistance on Social Dynamics}



\author{Zhe Liu}
\email{zheliu92@cs.ubc.ca}
\orcid{0000-0002-1904-9045}
\affiliation{%
  \institution{University of British Columbia}
  \city{Vancouver}
  \state{British Columbia}
  \country{Canada}
}

\author{Jiamin Dai}
\email{jdai24@cs.ubc.ca}
\orcid{0000-0002-4934-2760}
\affiliation{%
  \institution{University of British Columbia}
  \city{Vancouver}
  \state{British Columbia}
  \country{Canada}
}

\author{Joanna McGrenere}
\email{joanna@cs.ubc.ca}
\orcid{0000-0002-8411-3293}
\affiliation{%
  \institution{University of British Columbia}
  \city{Vancouver}
  \state{British Columbia}
  \country{Canada}
}

\renewcommand{\shortauthors}{Liu et al.}


\begin{abstract}
Eliciting rich data in semi-structured interviews is cognitively demanding, prompting recent work to explore real-time AI assistance for interviewers. However, introducing AI into the interviewer–interviewee interaction creates a triadic context whose social dynamics remain underexplored. We investigated how interviewers experience AI assistance for probing during semi-structured interviews. To elicit rich participant reflections, we implemented two variants of AI assistance differing in initiation and granularity in a high-fidelity prototype, \textit{ProbeAssist}. We conducted a qualitative-first comparative structured observation study where 18 participants each completed three simulated interviews: one without AI and two with different AI variants. Findings showed that participants leveraged AI as a supportive tool but resisted it as an assessor or competitor. As they navigated AI’s benefits and interaction costs, tensions emerged around agency, ownership, creativity, and interpersonal communication. We propose three implications for AI-assisted human-to-human interaction: managing social pressure, balancing idea alignment with inspiration, and preserving interpersonal presence.
\end{abstract}

\ccsdesc[500]{Human-centered computing~Interactive systems and tools}
\ccsdesc[500]{Human-centered computing~Collaborative and social computing design and evaluation methods}

\keywords{human-AI collaboration, AI, interview, qualitative research, large language model}


\maketitle

\section{Introduction}
Human-AI collaboration is rapidly expanding into complex professional domains \cite{wang2019human, jones2018searching, cai2019hello}, including qualitative research \cite{feuston2021putting}. To date, most research leveraging AI for qualitative methods has focused on supporting post-hoc analysis, such as coding \cite{gao2024collabcoder} and theme generation \cite{jiang2021supporting}, which occurs after data collection. Yet, the data collection stage is also critical, as the richness of the data captured during interviews directly shapes subsequent analyses \cite{golafshani2003understanding}. Semi-structured interviews, commonly used for qualitative data collection in fields including HCI, offer a flexible yet systematic approach to eliciting rich data \cite{roulston2010reflective}. This flexibility, however, places heavy demands on interviewers, who must simultaneously interpret interviewees’ responses, formulate follow-up questions, and manage conversational flow \cite{morris2015practical,roulston2010reflective}. Less experienced interviewers often struggle to know when and how to probe further \cite{peredaryenko2013calibrating}, while even experienced interviewers find navigating interviews cognitively demanding \cite{liu2025envisioning}, underscoring the opportunity for AI assistance. The growing use of video-mediated online interviews \cite{keen2022challenge} further creates space for integrating real-time AI assistance into interviews.

Current large language models (LLMs) can generate relevant follow-up questions and maintain conversational interactions \cite{deng2023rethinking}, raising the possibility of using AI to conduct interviews with limited human involvement. However, LLMs cannot fully capture the human empathy and contextual sensitivity required in qualitative interviewing \cite{montemayor2022principle}. Moreover, qualitative researchers value their agency \cite{roulston2010reflective} and have expressed concerns about approaches that may compromise the interpersonal connection central to interviewing \cite{liu2025envisioning}. These limitations suggest that rather than replacing human interviewers with AI, a promising direction is a human-in-the-loop approach \cite{wu2022survey}, in which AI supports interviewers while preserving their agency.

However, the real-time nature of interviews poses challenges for integrating AI capabilities into interaction design. Recent work has shown that AI can reduce interviewers’ cognitive load by supporting the combination of question formulation, note-taking, time tracking, and interview-script navigation \cite{wen2026interflow}. Yet, AI-generated follow-up questions might raise ethical concerns around unpredictable harms, unclear responsibility, and privacy and disclosure \cite{zhang2026ethics}. While these studies evaluated AI as a task-support tool and discussed the risks of AI-generated probing, less is understood about how interviewers themselves negotiate AI assistance within ongoing interview dynamics. We therefore investigate how interviewers experience and navigate the social and professional tensions that arise when AI is introduced into the interview, forming a triadic dynamic among interviewer, interviewee and AI.

This research seeks to advance the design of real-time AI assistance informed by the perspective of interviewers, guided by the following research questions (RQs):
\begin{itemize}
\item [RQ1:] How do interviewers engage with real-time AI assistance for probing in semi-structured interviews, compared with interviewing without AI assistance?
\item [RQ2:] How do interviewers perceive and navigate the impact of AI assistance in the interviewer--interviewee--AI triad?
\end{itemize}

Given the exploratory nature of these research questions, we adopted comparative structured observation \cite{mackay2025comparative}: a qualitative-first methodology that foregrounds participants’ experiences across design variants to deepen understanding and advance design, rather than evaluating particular prototypes. 

To provide diverse AI experiences to deepen participants' reflections, we drew on two design factors relevant to real-time AI assistance: \textit{initiation} \cite{allen1999mixed}, concerning whether assistance is provided on demand or proactively, and \textit{assistance granularity} \cite{desolda2025understanding}, representing the level of detail in AI suggestions, from brief keywords to fully formulated questions. We designed and implemented two variants of AI assistance within \textit{ProbeAssist}, a fully functional prototype for real-time AI assistance in online semi-structured interviews: (1) a restrained variant, which delivers assistance on demand with brief keywords, and (2) an expressive variant, which proactively suggests fully formulated follow-up questions. These variants enabled participants to experience diverse forms of AI-assisted interviewing and compare them with interviewing without AI assistance. We invited 18 participants with interviewing experience to each conduct three online simulated interviews---one without AI assistance and two with the AI variants---and share their experiences.

This paper makes three key contributions that advance the design of real-time AI interview assistance:
\begin{enumerate}
\item We provide empirical insights into how interviewers experience and navigate real-time AI assistance compared with unassisted interviewing, based on a qualitative-first study, showing that participants valued AI assistance for probing, yet their preferences for it were diverse and context-dependent.
\item We present the design and implementation of \textit{ProbeAssist}, a high-fidelity, open-source prototype for real-time AI assistance in online semi-structured interviews.
\item We identify emerging tensions around collaboration boundaries, agency, creativity, ownership, and the cognitive and interactional costs of AI, from which we derive three design implications for AI-assisted human-to-human interaction in professional contexts: mitigating social pressure, balancing idea alignment with novel inspiration, and preserving interviewers’ interpersonal presence with interviewees.
\end{enumerate}

\section{Related Work}
This section situates our study within emerging uses of AI in qualitative research and the broader human-AI collaboration literature, drawing on research across dyadic and triadic contexts to establish a foundation for understanding how AI’s involvement in the interview shapes the interviewer--AI collaboration.

\subsection{AI Advances for Qualitative Research}
\subsubsection{AI for Qualitative Study Design and Analysis}
AI is increasingly used to scaffold qualitative research across study design and analysis. In study design, LLMs have been leveraged for drafting protocols \cite{kamerlin2025using} and persona-based piloting \cite{sabbaghan2024role}. Despite their coherence, AI-generated accounts often lack contextual depth, while their human-like communication may misleadingly suggest genuine human experience, raising concerns about using them as proxies for human participants \cite{kapania2025simulacrum}. AI has also been incorporated into post-hoc qualitative analysis through systems such as PaTAT \cite{gebreegziabher2023patat}, and CollabCoder \cite{gao2024collabcoder} for collaborative coding. However, evaluations have identified risks of misinterpretation, over-trust \cite{bano2023ai}, and researcher deskilling \cite{schroeder2025large}. More broadly, most of these systems assume that researchers have sufficient time for deliberation \cite{morgan2023exploring}, limiting their suitability for the dynamic demands of real-time interviewing.

\subsubsection{Synchronous AI support during data collection}

AI has shown promise in supporting qualitative data collection, particularly in structured settings where conversational ambiguity is limited. For example, AI has been applied to surveys \cite{xiao2021let} and protocol-driven domains \cite{sun2025comparing}, achieving data quality comparable to human interviewers. However, qualitative interviews do not require strictly structured interactions, allowing interviewers to interpret responses and follow the natural flow of conversation. This flexibility can contribute to qualitative richness \cite{roulston2010reflective}, but challenges AI systems to adapt to conversations rather than enforce rigid protocols.

Researchers have explored AI as a fully autonomous interviewer in more flexible contexts, including commercial recruitment interviews \cite{aly2025optimizing}, motivated by the potential for scalability and consistency \cite{hu2024designing}. However, this approach faces important methodological limitations and conflicts with widely held qualitative research values \cite{schroeder2025large, liu2025envisioning}. Even with technical advances, AI agents cannot replicate the empathy, contextual sensitivity, or rapport-building required to ethically elicit rich qualitative data \cite{montemayor2022principle}. These limitations underscore the need for human interviewers to remain central to the process \cite{wu2022survey}, shifting the focus from scaling interviews toward supporting interviewers.

Consequently, researchers have begun to envision real-time AI assistants that support interviewers by flagging unasked questions or suggesting probes \cite{liu2025envisioning}. Yet, providing such assistance during a live interview remains challenging. LLM processing delays can render suggestions obsolete or disrupt conversational rhythm \cite{liu2025envisioning}, while interviewers must attend to the interviewee, maintain conversational flow, and decide whether and how to act on AI assistance. Recent work illustrates both the potential and limitations of AI-assisted interviewing. InterFlow \cite{wen2026interflow} showed that AI interview assistance can reduce interviewers’ cognitive load by supporting the combination of note-taking, time tracking, and interview script navigation, while also revealing limitations in the actionability of proactive suggestions; Zhang et al.\ \cite{zhang2026ethics} identified ethical concerns surrounding AI-generated follow-up questions in a Wizard-of-Oz interview setting, including the risk of harmful language and ambiguity in distributing responsibility between the interviewer and the AI.

Together, these studies demonstrate the technical feasibility of real-time AI support while highlighting important cognitive and ethical challenges. This motivates a closer examination of AI assistance from the interviewer’s perspective, particularly as it becomes part of an ongoing interviewer–interviewee interaction.

\subsection{Mechanisms and Dynamics of Human-AI Collaboration}
\subsubsection{Design factors in human-AI collaboration}
\label{sec:collab_mechanisms}

Mixed-initiative interaction \cite{allen1999mixed}, where control shifts between human and AI based on context \cite{horvitz1999principles}, is a design factor concerning how AI enters the human-AI interaction \cite{usmani2023human}. \textit{On-demand} interaction emphasizes user control \cite{holter2024deconstructing}, while \textit{proactive} AI provides timely support but may be intrusive \cite{yang2019unremarkable} or affect users’ sense of ownership \cite{shneiderman2020human}. In interviewing, these forms of initiative may offer different trade-offs between interviewer control and the convenience of timely assistance.

Another design factor is the granularity at which AI presents its suggestions, ranging from brief keywords for \textit{humans to interpret} to \textit{AI-formulated} output for direct use \cite{desolda2025understanding}. More granular suggestions can reduce user effort, whereas less granular suggestions leave more room for human interpretation and formulation. In the context of interviewing, InterFlow \cite{wen2026interflow} addressed this dimension by providing guidance that highlighted points warranting further exploration, rather than directly generating follow-up questions. Together, initiation and assistance granularity provide complementary ways to create diverse experiences of real-time AI assistance.

How AI contributes, however, is only part of the challenge. In human-AI collaboration, users must also be able to assess and negotiate AI contributions. Trust is therefore an important factor \cite{amershi2019guidelines}. Transparency and explainability have been shown to build trust by helping users assess AI outputs and make informed decisions about approving, modifying, or dismissing suggestions, supporting user control and satisfaction \cite{wu2022survey, gebreegziabher2023patat}. However, the fast-paced nature of interviews leaves interviewers little time to process complex explanations \cite{wuttke-etal-2025-ai}, creating a tension between transparency that supports trust and the cognitive cost of processing additional information.

This negotiation also relates to ownership, particularly in creative forms of human-AI collaboration \cite{biermann2022writers, wan2024felt}. Studies have found that substantial AI contributions can raise concerns about users’ ownership of and responsibility for the resulting work \cite{ohagi2024polarization}, whereas retaining users' control over AI contributions can help preserve their sense of authorship \cite{louie2020novice}. In semi-structured interviews, ownership is less straightforward because the interview is primarily a research activity for collecting qualitative data \cite{roulston2010reflective}, rather than creating an individual artifact. Nevertheless, qualitative researchers value the flexibility and serendipity of interviews \cite{feuston2021putting}, where interviewers’ interpretations and responses to unexpected directions can shape the data collected \cite{roulston2010reflective}. This raises an important question about how ownership applies when AI influences interviewers’ probing decisions rather than contributing to a discrete artifact.

\subsubsection{Social dynamics of human-AI collaboration}

Prior work on \textit{dyadic} human-AI collaboration shows that combining human intuition with AI’s analytical capabilities can improve task outcomes \cite{blanchard2024collective}, including higher accuracy than either method alone in medical diagnosis \cite{zoller2025human} and higher-quality co-created work in creative domains such as writing and music \cite{biermann2022writers, louie2020novice}. However, in many real-world settings, AI is introduced into ongoing human--human interactions rather than interacting with a single human, forming a \textit{triadic} system \cite{huang2025toward}. In these settings, AI can become part of the social interaction, influencing authority, communication, and trust \cite{castelfranchi1998modelling}. The doctor-patient-AI triad is a well-studied example \cite{lorenzini2023artificial}, where AI supports tasks such as documentation \cite{kocaballi2020envisioning}, prescribing \cite{leung2025ai}, and medical decision-making \cite{cai2019hello}, changing physicians’ roles as they interpret and communicate AI recommendations to patients \cite{lorenzini2023artificial}. Similar changes have also been examined in education (instructor-learner-AI) \cite{han2024teachers} and recruitment (recruiter-candidate-AI) \cite{aly2025optimizing}.

Integrating AI into semi-structured interviews similarly creates a triadic interaction, but its social dynamics differ from many existing triadic contexts. In clinical and educational settings, for example, triadic collaboration often assumes that both human parties are aware of the AI’s involvement \cite{han2024teachers, cai2019hello}. In contrast, research on AI-assisted interviewing suggests that interviewers differ in whether AI involvement should be disclosed to interviewees: some consider disclosure important, while others worry that revealing AI’s presence could alter the interaction and influence the qualitative data collected \cite{liu2025envisioning}. This highlights a distinct social consideration in interviews: AI may affect not only interviewer--AI collaboration but also the existing interviewer--interviewee interaction.

To examine these social dynamics, we draw on theoretical lenses that describe how people perceive and interact with AI as a social entity. The Computers Are Social Actors (CASA) paradigm, proposed by \citet{nass1994computers}, suggests that people unconsciously apply social rules from human interaction to technology. \citet{gambino2020building} extended this perspective to AI by proposing “human-media social scripts” that help explain why people may interact with AI differently from how they interact with other people. \citet{lee2010trust} further argued that users can perceive AI as a psychologically “real” entity, which can shape their engagement and trust \cite{pitardi2021alexa}. These perspectives provide a basis for examining how interviewers perceive and negotiate the AI’s role within the interviewer--interviewee--AI triad, especially because AI is introduced into an ongoing interpersonal interaction.

In sum, prior work has primarily positioned AI to support study design, post-hoc qualitative analysis, and data collection in highly structured settings, with emerging work extending AI assistance to real-time semi-structured interviews. However, how such assistance should be designed for the flexible and socially situated demands of semi-structured interviews remains underexplored. In particular, it remains unclear how interviewers experience AI-assisted interviewing compared with unassisted interviewing, how they navigate tensions around agency and ownership, and how they perceive the AI’s social role within the interviewer--interviewee--AI triad. To surface these experiences, we designed two variants of AI assistance informed by initiation and assistance granularity, enabling participants to reflect on diverse AI experiences.

\section{Designing \textit{ProbeAssist}}
We designed \textit{ProbeAssist}, a high-fidelity prototype to investigate the design of real-time AI assistance for semi-structured interviews. \textit{ProbeAssist} augments the interviewer interface with an AI assistance component, leaving the interviewee interface as a regular video call. The prototype design was guided by four design considerations derived from literature on qualitative methods and human-AI collaboration. These considerations were translated into concrete features through a user-centered design process \cite{abras2004user} involving brainstorming and design critiques. We distilled and implemented the features in two variants of AI assistance, \textit{restrained} and \textit{expressive}.

\subsection{Design Considerations}
By reviewing the challenges in semi-structured interviewing and the principles of effective human-AI collaboration, we established four high-level design considerations (DC1--DC4) to guide our exploration.

\begin{itemize}[leftmargin=1em]
\item \textbf{DC1: Support in-the-moment probing.} The prototype should provide timely, actionable assistance \cite{kontogianni2020tell} to help interviewers identify opportunities for deeper inquiry and enhance data richness \cite{morris2015practical, roulston2010reflective}.
\item \textbf{DC2: Deliver contextually relevant assistance.} The prototype should understand the research objectives and the interview context \cite{robinson2023probing, morris2015practical} to provide contextual assistance \cite{leverentz2025contextual}, as semi-structured interviews can take diverse and sometimes  unpredictable paths\cite{kostovicova2022harm}.
\item \textbf{DC3: Avoid cognitive and information overload.} The AI assistance should be glanceable \cite{yang2019unremarkable} and digestible \cite{schneider1987information}, integrating smoothly into the interviewer's workflow \cite{wang2025less} without distracting \cite{hemmer2023human} or causing cognitive and information overload \cite{fox2007available}.
\item \textbf{DC4: Preserve interviewer control.} Recognizing that qualitative researchers value their professional judgment \cite{feuston2021putting, jiang2021supporting} and resist AI systems that undermine their agency \cite{liu2025envisioning}, the prototype should support interviewers while ensuring they retain primary control \cite{Fgener2021CognitiveCI}.
\end{itemize}

\subsection{Prototype Design}
To translate these design considerations into a prototype, we conducted a two-hour brainstorming session with 7 HCI researchers, designers, and qualitative researchers from our institution. The session generated candidate features for real-time AI assistance, which we refined into low-fidelity sketches and discussed in a subsequent design critique session with 30 HCI researchers and students. Participants evaluated the features' usefulness, potential intrusiveness, and alignment with interview practice. This feedback helped prioritize features for further exploration rather than converging on a single optimal design. We then selected, clustered, and merged these features into two variants of AI assistance. Guided by the four design considerations described above, we further drew on initiation and assistance granularity to differentiate the variants, providing participants with substantively different AI experiences for reflection.

\subsubsection{Core Features}
Our design is grounded in three core features adhering to the design considerations:

\begin{enumerate}
\item \textbf{Dual-Panel Display:} The interface is spatially divided into a video call panel (left) and an AI assistance panel (right), enabling interviewers to stay present with interviewees while accessing contextual AI assistance (\textbf{DC2}). This feature aims to keep AI assistance accessible yet unobtrusive, thereby helping interviewers concentrate their cognitive resources (\textbf{DC3}) and preserving their control (\textbf{DC4}).
\item \textbf{Initiative Modes:} The prototype offers two modes of initiation, \textit{on-demand} and \textit{proactive}, allowing participants to experience and reflect on the tension between interviewer control (\textbf{DC4}) and timely assistance (\textbf{DC1}).
\item \textbf{Contextual AI Assistance:} To support interviewers with probing (\textbf{DC1}), the prototype provides two types of real-time assistance grounded in the live transcript (\textbf{DC2}), reflecting the \textit{assistance granularity} dimension: \textit{keywords}, presented as a summary of the interviewee’s response along with several probing directions; and \textit{phrases}, which provide ready-to-use follow-up questions in full sentences. To address information load (\textbf{DC3}), both types present a similarly manageable amount of text.
\end{enumerate}

\begin{table}[ht]
\centering
\caption{Two variants of AI assistance, informed by initiation and assistance granularity.}
\label{tab:variant_space}
\Description{The table shows two AI assistance variants, restrained and expressive, across three dimensions: dual-panel display, initiation, and assistance granularity. Both variants use a dual-panel display. Restrained assistance is on-demand and provides keywords, whereas expressive assistance is proactive and provides phrases.}
\begin{tabular}{lccc}
\toprule
Variant & Dual-panel display & Initiation & Assistance granularity \\
\midrule
\textit{Restrained} & \checkmark & On-demand & Keywords \\
\textit{Expressive} & \checkmark & Proactive & Phrases \\
\bottomrule
\end{tabular}
\end{table}

\subsubsection{Two Variants of AI Assistance}

\begin{figure}[ht]
    \centering
    \includegraphics[width=\linewidth]{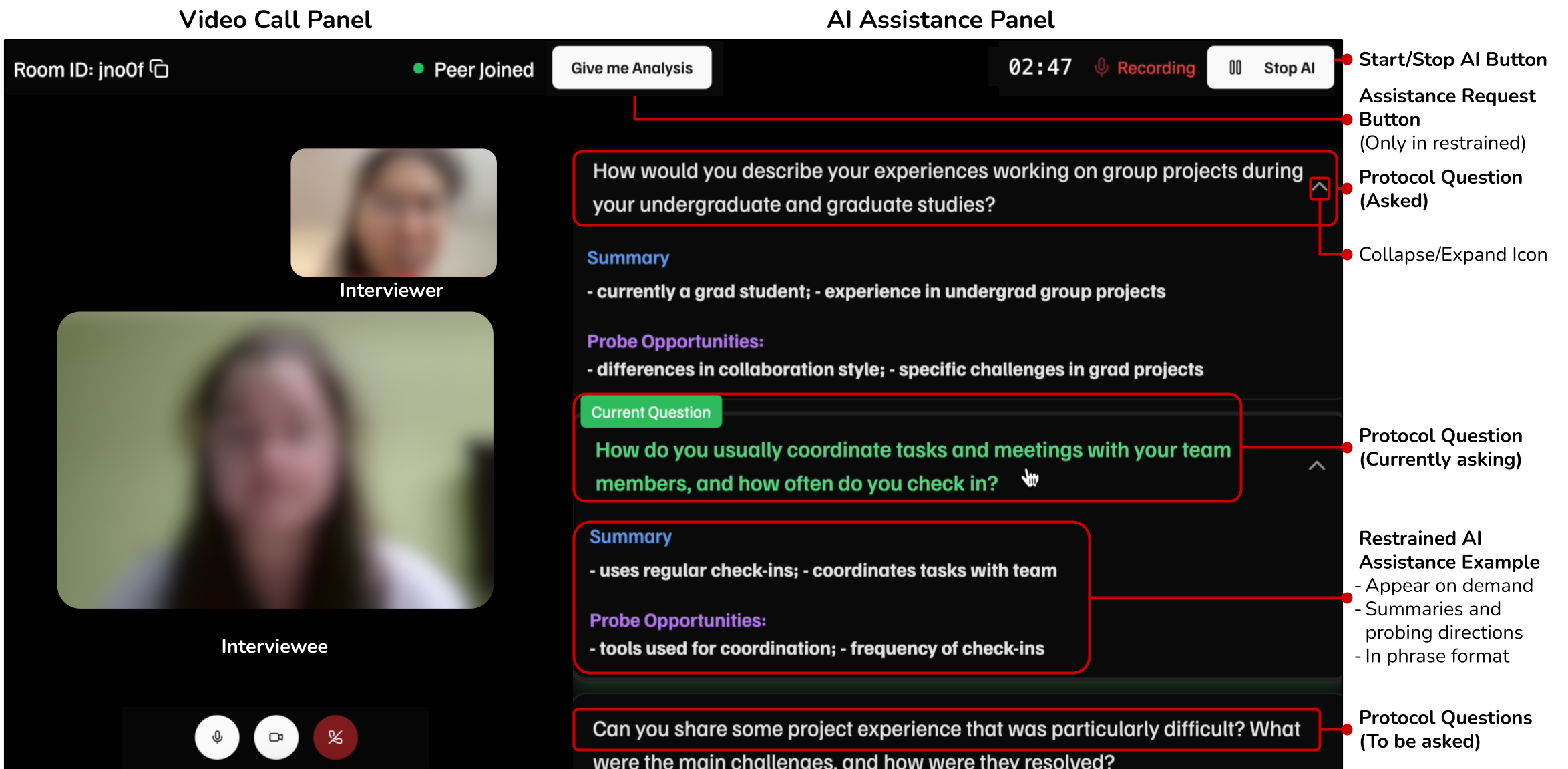}
    \caption{Annotated screenshot of the \textit{ProbeAssist} interface in the \textit{restrained} variant. The video call panel (left) is augmented with an AI assistance panel (right) that provides summaries and probe opportunities with brief keywords. After the \textit{Start AI} button is activated, the AI assistance appears only when the interviewer clicks the \textit{Give me Analysis} button each time. Interviewer and interviewee video captures are blurred for anonymity in publication.}
    \label{fig:ODGinterface}
    \Description{This figure shows the user interface with a dark background, divided into a smaller video call panel on the left and a larger AI assistance panel on the right. The video call panel contains a header with a room ID, two vertically stacked video feeds with blurred faces labeled 'Interviewer' (top, small) and 'Interviewee' (bottom, large), and icons for microphone, camera, and ending the call below the videos. The AI assistance panel displays a 'Give me Analysis' button for the interviewer to request assistance, a timer showing the interview time with a 'Recording' indicator, and a 'Stop AI' button, which the interviewer can use to start or stop the AI assistance, at the top. Below them is a vertical display of the interview questions and the corresponding AI assistance, showing a sequence of questions asked, currently being asked, and to be asked by the interviewer. Following each question-and-answer exchange, the panel presents an AI-generated summary of the interviewee's response and a list of potential probe opportunities in phrase format. For example, for the 'Current Question' ('How do you usually coordinate tasks and meetings…'), the AI has generated a summary ('uses regular check-ins…') and suggested probe opportunities ('tools used for coordination; frequency of check-ins'). There is a collapse/expand icon next to the AI assistance for user control. The annotations on the right side explain these elements, highlighting that the assistance request button is in the restrained variant only.}
\end{figure}

\begin{figure}[h]
    \centering
    \includegraphics[width=0.7\linewidth]{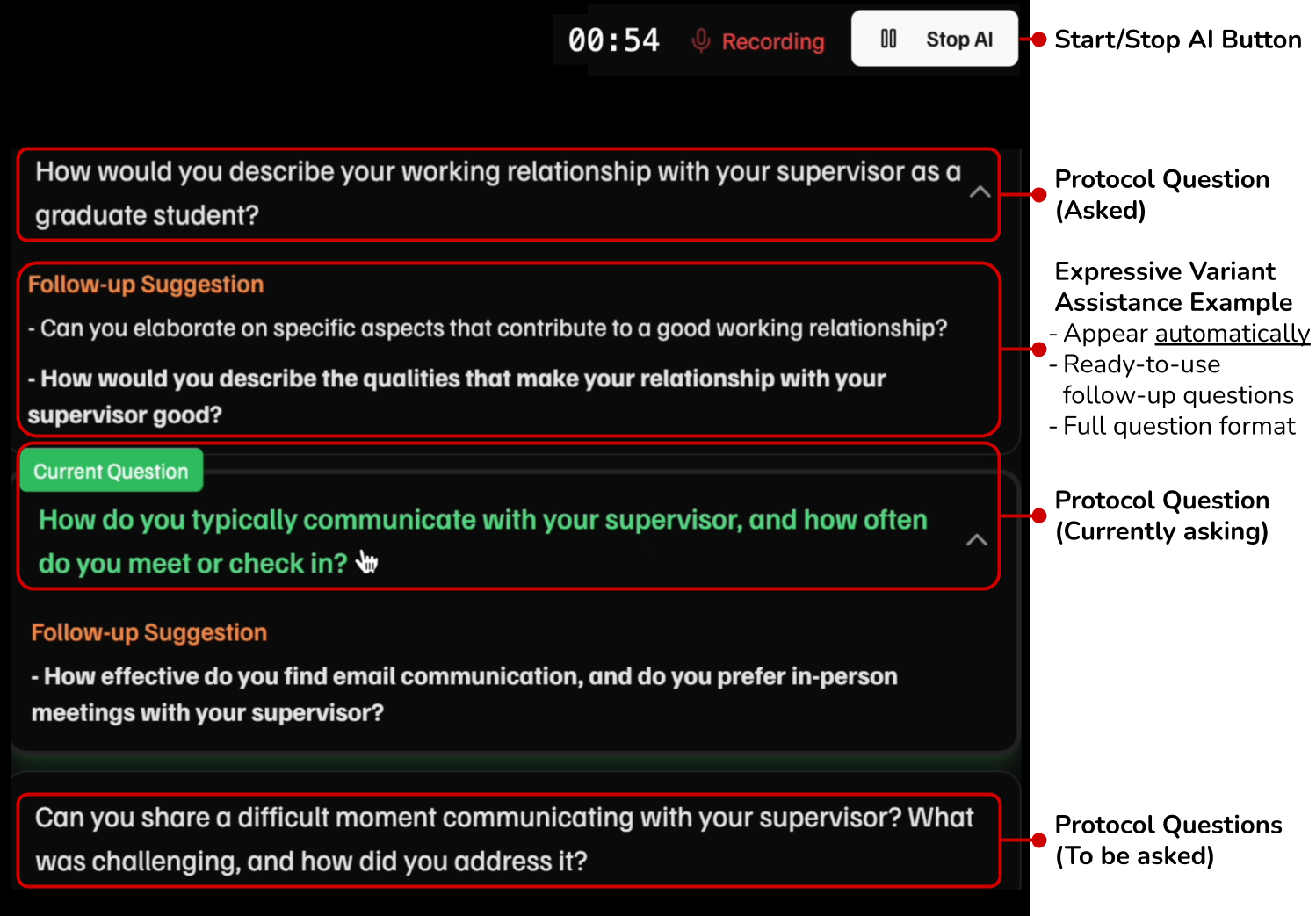}
    \caption{Annotated screenshot of AI assistance panel in the \textit{expressive} variant. After the \textit{Start AI} button is activated, follow-up questions appear automatically once the prototype detects the interviewee has finished answering. More details on both variants are in the video figure.}
    \label{fig:PSinterface}
    \Description{This figure shows a screenshot of an AI assistance panel with a dark background. At the top, a header shows a timer displaying the interview time, a 'Recording' indicator, and a 'Stop AI' button. Below them is a vertical display of the interview questions and the corresponding AI assistance, showing a sequence of questions asked, currently being asked, and to be asked by the interviewer. Following each question, the panel presents a list of AI-generated follow-up suggestions in a full-question format. For example, for the 'Current Question' ('How do you typically communicate with your supervisor...'), the AI has generated follow-up suggestions such as 'How effective do you find email communication, and do you prefer in-person meetings with your supervisor?'. The annotations on the right side explain these elements, highlighting that the assistance appears automatically in the expressive variant.}
\end{figure}

Drawing on these core features, we designed two variants of AI assistance to provide participants with diverse experiences for reflection (see Table~\ref{tab:variant_space}). The \textit{restrained} variant (see Fig.~\ref{fig:ODGinterface}) gives the interviewer control over \textit{when} to receive assistance and \textit{how} to interpret it, combining on-demand initiation with keyword-level assistance. The \textit{expressive} variant (see Fig.~\ref{fig:PSinterface}) proactively provides ready-to-use follow-up questions, combining proactive initiation with phrase-level suggestions. Together, these variants offered participants substantively different AI experiences---one emphasizing interviewer control and interpretation, the other emphasizing AI initiative and formulation---to ground and deepen their reflections on AI-assisted interviewing. Other configurations (e.g., on-demand with phrases, proactive with keywords) remain possible and could be explored in future work. A demonstration of the \textit{restrained} and \textit{expressive} variants is provided in our video figure.

\begin{figure}[ht]
    \centering
    \includegraphics[width=0.45\linewidth]{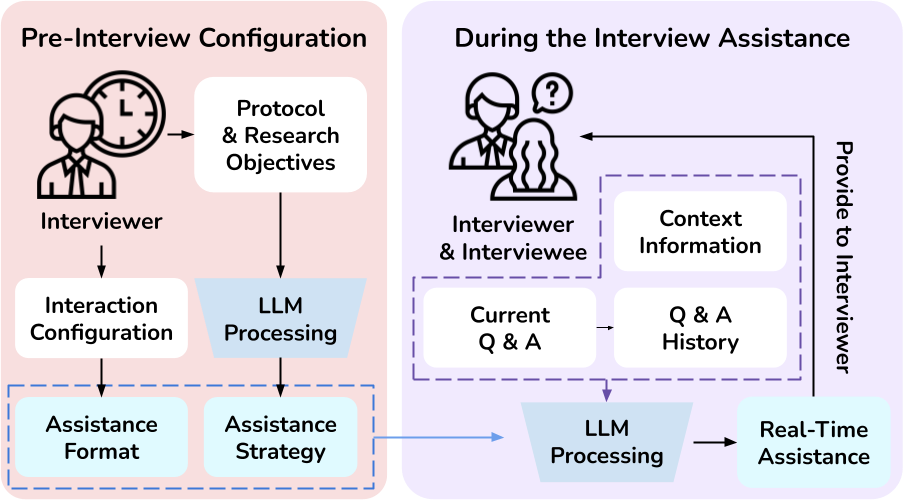}
    \caption{Interaction flow of the \textit{ProbeAssist}, illustrating how the interviewer configures the prototype before the interview and receives real-time, contextual AI assistance during the session.}
    \label{fig:interaction}
    \Description{This figure is a flowchart illustrating the prototype’s interaction flow, which is divided into two phases: pre-interview configuration (left panel) and during the interview assistance (right panel). In the pre-interview configuration phase, the interviewer provides two types of input: the interaction configuration and the protocol with research objectives. The protocol and research objectives are processed by a large language model. The output of this processing, combined with the interaction configuration, establishes the assistance strategy and format to be used during the interview. The assistance strategy and format then inform the assistance during the interview. In this phase, the current question and answer discussion from the conversation between the interviewer and interviewee is captured and used to build a Q and A history. This history, along with other context information, is fed into another LLM processing step. This step generates real-time assistance, which is then delivered back to the interviewer, completing the assistance loop.}
\end{figure}

\subsubsection{Interaction Flow}
The interaction flow aligns with the interviewer’s existing workflow (see Fig. \ref{fig:interaction}) and addresses our design considerations. Using the web-based prototype, the interviewer first uploads the interview guide and research objectives (\textbf{DC2}), as provided in Appendix \ref{appendix:participant_protocol}, and selects one variant (\textbf{DC4}) before initiating the call and sharing a link with the interviewee. Once the interviewee joins, the interviewer activates AI support before asking any questions. During the session, the prototype provides real-time assistance (\textbf{DC1}) in the AI assistance panel according to the chosen variant, allowing the interviewer to freely accept, adapt, or dismiss the assistance (\textbf{DC4}). After the interview, the interviewer deactivates AI support and ends the call. The prototype generates a downloadable package with the video recording, transcript, and AI usage log.

\subsection{Prototype Implementation}
\textit{ProbeAssist} is implemented as a high-fidelity, web-based interactive research prototype to ground participants’ reflections in their experience with actual real-time AI assistance. The front end is developed using Next.js 15 with React 19. Peer-to-peer video communication is established through WebRTC, while session management and signaling are handled by Firebase Firestore. The assistance pipeline captures and streams mixed audio using the Web Audio API and FFmpeg.wasm. We then leverage OpenAI’s Realtime API (\texttt{gpt-4o-realtime-preview}\footnote{The prototype was implemented in 2025, when OpenAI’s Realtime API was among the state-of-the-art APIs for real-time audio processing and interaction.}) for immediate audio transcription and analysis. By analyzing the transcripts against a structured JSON representation of the interview guide, the prototype generates and displays contextual assistance (prompts provided in Appendix \ref{appendix:prompt}). To support transparency and future work, we will release the prototype code as open source.

\section{Methods}
Given the exploratory nature of our research questions, we conducted an in-person comparative structured observation (CSO) study \cite{mackay2025comparative} to investigate interviewers' nuanced experiences with real-time AI assistance. To elicit richer reflections, participants experienced two variants of AI assistance alongside an unassisted experience. CSO is well-suited to this context because it adopts a qualitative-first approach that foregrounds participants' experience and reflective comparisons, enabling systematic exploration of how and why design should be advanced, rather than evaluating the performance of any specific prototype. Accordingly, we focused primarily on the qualitative analysis of participants' debriefing interviews to understand their perspectives, and to surface design tensions and opportunities surrounding AI assistance. Quantitative measures from surveys and logs were collected to complement the qualitative insights.

\subsection{Study Design}
\subsubsection{Simulated interviews}
To explore the nuances of our design variants under a consistent interactional context, we simulated an online interview setup to replicate interviewing experiences as realistically as possible. Three distinct semi-structured interview guides were prepared and provided by us on (1) time management, (2) team project experience, and (3) supervisory relationships, all similarly familiar to participants and not requiring specific domain expertise (see Appendix \ref{appendix:participant_protocol}). Each guide was carefully designed with a similar flow, scope, and depth to ensure the comparative value of our study. To simulate typical interview preparation, participants received the guides at least 24 hours before the session to prepare in their usual ways. During the study, participants conducted three online simulated interviews with one interviewee who joined the video call from another building. To minimize any feeling of being assessed, participants were left alone in the room during the interviews. Participants were given the impression that they were interviewing a real, specifically recruited interviewee, but the role of the interviewee was consistently performed by a trained research assistant. Participants were informed of this role-playing arrangement at the end of the study.

We used a single trained role-player to provide a relatively consistent interactional context across sessions and avoid interviewees being overly talkative or reserved \cite{wen2026interflow}. This allows us to focus on how participants experienced the AI assistance, as differences in interviewees' response style could otherwise influence their experience independently of the AI assistance. It also reduced the privacy and ethical risks of processing real interviewees’ sensitive information with an LLM in real time \cite{zhang2026ethics}. Participants nevertheless conducted actual interviews with a responsive interviewee, made real-time probing decisions, and managed conversational flow. 

To ensure fidelity and consistency, the research assistant acting as the interviewee followed a detailed role-playing protocol. She was instructed to familiarize herself with all interview topics while responding spontaneously during the interviews, grounding her answers in her actual experiences to maintain authenticity. She was also explicitly instructed to avoid providing exhaustive answers by default, leaving space for interviewers to probe further and thereby creating realistic opportunities to leverage AI assistance. Additionally, she was directed to adjust the depth of her responses based on the interviewer's approach, offering more elaborate answers when prompted effectively and reacting authentically to closed-ended or unclear questions (See Appendix \ref{appendix:role_play} for the full role-playing protocol).
\begin{figure}[h]
    \centering
    \includegraphics[width=\linewidth]{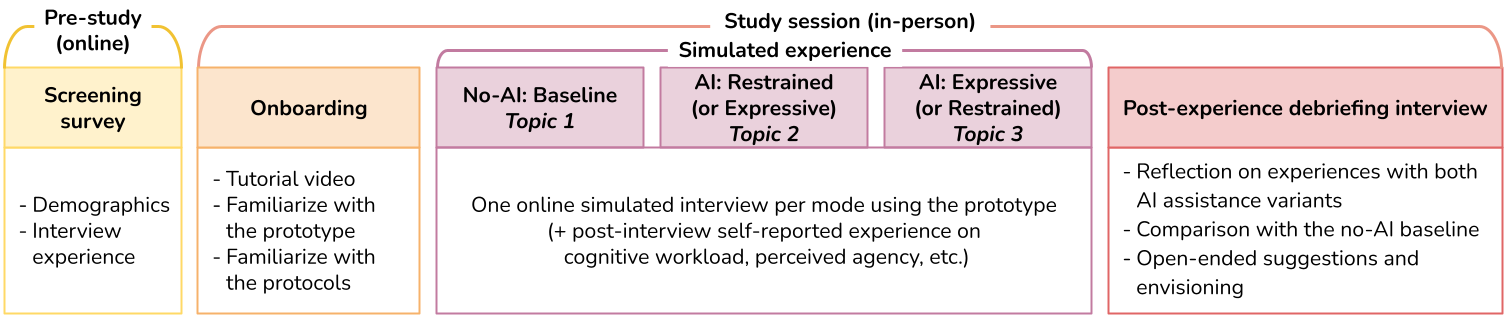}
    \setlength{\abovecaptionskip}{-5pt}
    \setlength{\belowcaptionskip}{-5pt}
    \caption{Overview of the study procedure}
    \label{fig:procedure}
    \Description{This figure is an overview of the study procedure, outlining an online pre-study phase and an in-person study session. The pre-study phase consists of a screening survey used to collect participant demographics and prior interviewing experience. The in-person study session begins with an onboarding period, during which participants watch a tutorial video and familiarize themselves with the prototype and interview guides. This is followed by the main study block, where participants conduct online simulated interviews on three topics under three different settings. Topic 1 is conducted without AI assistance. For Topics 2 and 3, participants use the two counterbalanced AI assistance variants: one topic with the restrained variant and the other with the expressive variant. After completing each interview, participants provide post-interview self-reported ratings on their cognitive load, perceived control, etc. The session concludes with a post-experience debriefing interview, during which participants reflect on their experiences with both AI assistance variants, compare them with interviewing without AI assistance, and answer open-ended questions for potential future use.}
\end{figure}

\subsubsection{Procedure}
Each study session, lasting approximately 90 minutes in person, was structured into three parts: (1) onboarding and preparation, (2) engagement with prototype variants through three online simulated interviews, and (3) a debriefing interview (see Fig. \ref{fig:procedure}). During the 15-minute onboarding, participants watched a video tutorial on the \textit{ProbeAssist} prototype. They were then given time to familiarize themselves with the prototype, review three interview guides, and prepare for the simulated interviews.

Following the onboarding, participants conducted three 10–15 minute online simulated interviews. The first interview was conducted without AI assistance, followed by the two variants of AI assistance (\textit{restrained} and \textit{expressive}). The unassisted interview was placed first as it most closely reflected participants’ typical online interview experience, providing a familiar warm-up and an unassisted reference for reflecting on the subsequent AI-assisted experiences. The order of the \textit{two AI variants} and the assignment of the \textit{three interview guides} were counterbalanced across participants. After each simulated interview, participants completed a short survey to assess their cognitive load (NASA-TLX), self-perceived interview quality, and their sense of control. The session concluded with a 30-minute semi-structured debriefing interview where the first author invited participants to reflect on their experience with AI-assisted interviewing compared with interviewing without AI assistance, focusing on their (1) overall experience, (2) perceived data richness, (3) cognitive load and sense of control, and (4) design requirements for future iterations. Participants were invited to use their concrete experience with the two variants as a starting point, then expand their reflections to discuss their broader preferences, concerns, and expectations for AI assistance in interviews generally, including scenarios and features beyond what the prototype offered. The full survey, debriefing protocol, and a video figure demonstrating all variants are available in the supplemental materials.

\begin{table}[htbp]
\small
\centering
\caption{Participant background}
\noindent
\begin{NiceTabular}{
    >{\centering}m{0.04\linewidth}%
    >{\centering}m{0.05\linewidth}%
    m{0.50\linewidth}%
    m{0.27\linewidth}%
    }[notes/para,
      code-before = \rowcolors{3}{tablegray}{}
      ]
\toprule
 \textbf{PID\tabularnote{Self-reported expertise in PID: n = novice, i = intermediate, e = expert}} & \textbf{Gender\tabularnote{W = woman, M = man, N = non-binary}} & \centering \textbf{Self-reported Interviewing Experience} & \centering\arraybackslash \textbf{Self-reported AI Experience}%
\tabularnewline%
\midrule%
P1e & W & Academic; Remote; 6-year experience in empirical studies with \textgreater 50 interviews, primarily in HCI and behavioral research & Knowledgeable about AI and frequently uses it at work\\
P2n & W & Academic; Remote; Experience in usability evaluation within one student-led project & Familiar with AI concepts and capabilities\\
P3i & W & Academic; In-person and remote; 4-year experience in career education, primarily focusing on descriptive rather than in-depth analysis; Highly experienced in job interview training & Finds AI helpful for work but worries about misuse\\
P4e & M & Academic; In-person and remote; 8-year experience in psychology and education research & Open to AI applications with careful design and testing\\
P5n & N & Industry; In-person; Conducted 2 user studies with a total of 18 participants; Not active since 2022 & Familiar with AI concepts but skeptical about AI usage at work\\
P6n & W & Non-profit and academic; In-person; Conducted 12 interviews on community engagement in social science & Knowledgeable about AI and frequently uses AI tools at work\\
P7n & M & Academic; In-person; Experience in HCI course projects with \textless 20 interviews & Frequently uses AI for both academic and daily tasks\\
P8e & M & Academic; Remote; 4-year experience in HCI with \textgreater 50 interviews, conducting rigorous qualitative research studies & Tried AI in research and plans to explore further\\
P9n & W & Academic; In-person and remote; 2-project experience in HCI and health care exploratory interviews & Familiar with AI concepts but resistant to using AI in daily life\\
P10n & W & Academic; In-person; 1-year experience in psychology and higher education projects & Has tried AI in daily tasks; Interested in applying it to academic work\\
P11i & W & Industry and academic; Remote; 3-year experience in HCI and psychology; Received rigorous training from supervisor & AI practitioner; Positive about using AI for research\\
P12e & M & Industry and academic; In-person and remote; 10-year experience in law and policy research & Recently started using AI tools and now finds them helpful\\
P13i & W & Industry and academic; In-person and remote; 4-year experience in HCI expert interviews, often using mixed-method approaches & AI practitioner; Researches AI for mental health\\
P14n & M & Academic; In-person; 1-year experience in communications research and small-scale discourse analysis & Finds AI helpful and efficient for writing\\
P15i & M & Non-profit and academic; In-person and remote; 4-year experience in social science and youth development & Rarely uses AI but keeps up with AI discussions\\
P16i & M & Academic; In-person and remote; 5-year experience in education and social support research & Curious about AI in daily life but only uses it occasionally\\
P17i & M & Academic; In-person and remote; 3-year experience in education and psychology; Received formal training in qualitative methods & Has tried AI in research and plans to explore it further\\
P18i & N & Academic; In-person and remote; 5-year experience in systems and affective computing, integrating technical and human evaluation & Uses AI at work and seeks to expand usage\\
\bottomrule
\end{NiceTabular}
\label{table:participant_info}
\end{table}
\Description{This table describes the backgrounds of the 18 participants, reporting their participant ID (PID), gender, self-reported interviewing experience, and self-reported AI experience. The PID includes a suffix for self-reported expertise: n = novice, i = intermediate, e = expert. Gender is abbreviated as W = woman, M = man, and N = non-binary.
P1e, Woman, 6 years of academic experience in HCI with >50 interviews; Knowledgeable and frequent AI user at work.
P2n, Woman, Academic experience from one student-led project; Familiar with AI concepts.
P3i, Woman, 4 years of interviewing experience in career education and >20 years in job interview training; Finds AI helpful but worries about misuse.
P4e, Man, 8 years of academic experience in psychology and education; Open to AI with careful design.
P5n, Non-binary, Industry experience with 2 user studies; Familiar with AI concepts but skeptical of its use at work.
P6n, Woman, Non-profit and academic experience with 12 interviews in social science; Knowledgeable and frequent AI user at work.
P7n, Man, Academic experience from HCI course projects with <20 interviews; Frequent AI user for academic and daily tasks.
P8e, Man, 4 years of academic experience in HCI with >50 interviews; Has tried AI in research and plans to explore further.
P9n, Woman, Academic experience from 2 projects in HCI and healthcare; Familiar with AI concepts but resistant to using it in daily life.
P10n, Woman, 1 year of academic experience in psychology and higher education; Has tried AI for daily tasks and is interested in academic applications.
P11i, Woman, 3 years of industry and academic experience in HCI and psychology; AI practitioner, positive about AI for research.
P12e, Man, 10 years of industry and academic experience in law and policy; Recently started using AI tools and finds them helpful.
P13i, Woman, 4 years industry and academic experience in HCI expert interviews; AI practitioner who researches AI for mental health.
P14n, Man, 1 year of academic experience in communications research; Finds AI helpful and efficient for writing.
P15i, Man, 4 years of non-profit and academic experience in social science; Rarely uses AI but follows AI discussions.
P16i, Man, 5 years of academic experience in education and social support; Curious about AI but uses it only occasionally.
P17i, Man, 3 years of academic experience in education and psychology; Has tried AI in research and plans to explore it further.
P18i, Non-binary, 5 years of academic experience in systems and affective computing; Uses AI at work and seeks to expand usage.}

\subsection{Participants}

To gather high-quality design feedback, after obtaining ethics approval from our institution (a large public university in North America), we recruited 18 participants both from the university, via departmental mailing lists, and from the broader qualitative research community through online platforms, supplemented by snowball sampling. The sample size was chosen to align with the local standards for qualitative HCI studies, aligning with Caine’s guidelines \cite{caine2016local} and recent human-AI collaboration work \cite{feuston2021putting, jiang2021supporting}. The inclusion criteria required participants to be fluent in English and to have prior experience conducting semi-structured interviews, ensuring they could meaningfully reflect on the design tensions from an interviewer’s perspective. All interested individuals completed a screening questionnaire to confirm their eligibility before being invited to the study. While we did not aim for full representativeness of such a broad population, the final sample reflected diversity in gender, research background, interviewing experience and expertise, and AI experience. Participants’ self-reported expertise levels are noted in their participant IDs (PIDs) to provide context for their quotes.

The 18 participants (see Table \ref{table:participant_info}) consisted of 8 women, 8 men, and 2 non-binary individuals from Computer Science, Information, Psychology, Health Care, Education, and Sociology. Participants have 1–10 years of interview experience (averaging 3.6 years), mostly in academic settings, with several also experienced in industry research interviews. Each participant was compensated with \$35.

\subsection{Data Collection and Analysis}
Aiming to advance AI assistance design, we prioritized qualitative analysis of debriefing interviews to capture the nuances of participants’ reflections. The interviews (35–64 minutes each) were audio-recorded, transcribed using a local AI-powered speech recognition tool, and manually cleaned and proofread. We conducted an inductive reflexive thematic analysis \cite{braun2021thematic} of all transcripts. The first author open-coded the transcripts in NVivo 12 after data collection was completed, regularly discussing emerging codes and themes with co-authors to ensure rigor and consistency.

To complement and contextualize our primary qualitative analysis, we collected supporting quantitative data from interview recordings and post-interview surveys. The recordings captured screen activity and conversation audio, from which we derived behavioral metrics of participants' interactions with the variants. Post-interview surveys included the NASA-TLX for cognitive load \cite{hart1988development}, custom items assessing perceived interview quality \cite{robinson2023probing}, perceived control \cite{holter2024deconstructing, gomez2023mitigating} and the impact of AI assistance \cite{fragiadakis2024evaluating}. All custom items used a 7-point Likert scale \cite{joshi2015likert}. 

\subsection{Positionality Statement}
We reflect on our positionality and its potential influence on the research process. The first author, as the lead researcher (early 30s, Computer Science), has practical interviewing experience in both academic and industry settings and values semi-structured interviews as a complex and meaningful method. With a generally positive outlook on AI technologies, they see human-AI collaborative interviewing as an inevitable evolution. The two co-authors bring complementary perspectives: one (Information Studies) is a qualitative researcher with extensive experience conducting semi-structured interviews in sensitive contexts; the other (Computer Science) has substantial experience in both quantitative and qualitative research. Collectively, all authors acknowledge that AI development is reshaping research practices, recognizing the urgency of examining human-AI collaboration in this context. This diversity of backgrounds integrates multiple disciplinary perspectives, strengthening our study design and data analysis.



\section{Findings}
\label{findings}

In this section, we first report how participants used and perceived AI assistance compared with unassisted interviewing, including their reflections on the two AI variants (\S\ref{sec:usage_perception}). We then present three themes from our thematic analysis of debriefing interviews, examining how participants structured the AI's role (\S\ref{theme1_role_foundation}), balanced its benefits against cognitive and social costs (\S\ref{theme2_benefit_cost}), and navigated tensions around agency, creativity, and ownership (\S\ref{theme3_tension}).


\subsection{Usage, Perceptions, and Preferences}
\label{sec:usage_preception_preference}

\subsubsection{Usage and perceptions}
\label{sec:usage_perception}

Quantitative results contextualize how participants used and evaluated the AI assistance. Here, $N$ denotes interviews without AI assistance, $R$ and $E$ denote the \textit{restrained} and \textit{expressive} variants, respectively (e.g., $M_R$, $p_R$). We found that participants actively incorporated the assistance into their natural probing process. Usage data (Table \ref{tab:followup}) shows that participants asked significantly more total follow-up questions with both AI variants than without AI assistance ($M_R=9.83, M_E=8.72$ versus $M_N=5.83$, $p=.005$ for both pairwise comparisons, Tukey’s HSD following a significant one-way ANOVA). Importantly, this increase appeared to be additive rather than substitutive: participants incorporated or adapted approximately three \textit{AI-assisted} questions per session ($M_R=3.39, M_E=3.00$) in addition to a similar number of \textit{self-formulated} questions ($M_R=6.44,\ M_E=5.72$) as without AI assistance ($M_{N}=5.83$).
Not surprisingly, given the significant difference in number of total questions, pairwise comparisons show that restrained ($M_R = 13.76$ min) and expressive ($M_E = 12.65$ min) variants resulted in significantly longer interviews than without AI assistance ($M_N = 10.12$ min; $p_R <.001, p_E =.005$); the difference between AI variants was not significant ($p=.134$).

Participants also expressed positive experiences with both AI variants, with AI impact ratings above 5.14 out of 7. Their perceived interview quality was significantly higher with both AI variants ($M_R=6.04,\ M_E=5.90$) than without AI assistance ($M_{N}=5.56,\ p_R=.006,\ p_E=.042$; Tukey’s HSD and one-way ANOVA). They likewise reported a high sense of control with both AI variants ($M_R=6.28,\ M_E=6.21$), while their self-reported cognitive load did not significantly differ across the three simulated interviews, as shown in Appendix \ref{visualization}.

\begin{table}[ht]
\centering
\begin{threeparttable}
\caption{Average number of follow-up questions across interviews.}
\begin{tabular}{lccc}
\hline
\textbf{Interviews} & \textbf{Total Follow-up Qs} & \textbf{AI-Assisted Follow-up Qs} & \textbf{Self-Formulated Follow-up Qs} \\
\hline
No-AI  & 5.83 & - & 5.83 \\
Restrained  & $$ 9.83\text{*} $$ & 3.39 & 6.44 \\
Expressive  & $$ 8.72\text{*} $$ & 3.00 & 5.72 \\
\hline
\end{tabular}
\begin{tablenotes}
\item * indicates a significant difference ($p<0.01$) from the unassisted experience.
\end{tablenotes}
\label{tab:followup}
\Description{This table presents the average number of follow-up questions asked in an unassisted interview and in interviews using the restrained and expressive AI assistance variants. The data are organized into total follow-up questions, AI-assisted follow-up questions, and self-formulated follow-up questions. In the unassisted interview, participants asked an average of 5.83 follow-up questions, all of which were self-formulated. With the restrained variant, participants asked an average of 9.83 follow-up questions, including 3.39 AI-assisted and 6.44 self-formulated questions. With the expressive variant, participants asked an average of 8.72 follow-up questions, including 3.00 AI-assisted and 5.72 self-formulated questions. An asterisk indicates that the total number of follow-up questions in both AI variants is significantly different from the unassisted interview.}
\end{threeparttable}
\end{table}

\subsubsection{Preferences}
\label{variant_reflections}

Across both AI variants, participants described the assistance as useful for reassuring their probing direction and surfacing novel probing opportunities. They also recognized the need for interviewers to evaluate and filter suggestions based on the interview context.

Regarding initiation, participants weighed the value of on-demand versus proactive assistance based on their in-the-moment needs for control or convenience during the interview. Participants who preferred on-demand initiation prioritized control, noting that they “would not be disturbed” (P1e) and could “decide when to receive assistance” (P16i). In contrast, participants who preferred proactive assistance saw requesting assistance as “extra work” that could be “distracting” (P9n), noting that “when [the assistance] just popped up, it’s way more natural” (P6n). These preferences could also vary within the same participant: P11i preferred to retain control when she had “a clear line of inquiry,” but wanted proactive assistance when she felt “scattered.” 

Regarding assistance granularity, participants similarly weighed the flexibility of keywords against the lower formulation effort of ready-to-use phrases. Participants who preferred the form of keywords appreciated having the “flexibility” (P13i) to “formulate questions in [their] own words” (P2n) and maintain their “personal style” (P17i). In contrast, participants who preferred the expressive variant found the ready-to-use output “simpler to process” (P11i) and noted that they did not “have to pause to construct a sentence” (P9n). Interestingly, these preferences were not always mutually exclusive: P6n proposed combining the two forms by presenting the full question with “[bolded] keywords,” allowing her to “quickly decide […] whether to read the full question.” 



\vspace{1.5em}
\noindent Our thematic analysis examines the deeper motivations, trade-offs, and tensions underlying participants’ experiences with AI assistance. Although some reflections were prompted by features specific to one variant, participants often extended these reflections to AI assistance more broadly. We present the thematic findings through three interconnected themes (see Table \ref{theme_table}). The first theme (\S\ref{theme1_role_foundation}) describes how participants structured the AI’s role and established collaboration boundaries. The second theme (\S\ref{theme2_benefit_cost}) highlights practical trade-offs in using \textit{ProbeAssist}, balancing the functional benefits of AI assistance against the cognitive and social costs of integrating AI into live interviews. Building on these, the third theme (\S\ref{theme3_tension}) surfaces deeper tensions around agency, creativity, and professional ownership.

\begin{table}[ht]
\caption{Summary of themes/subthemes, highlights, and their relation to the research questions}
\centering
\begin{tabular}{m{0.55\linewidth}m{0.41\linewidth}}
\hline
\textbf{Theme/subthemes} & \textbf{Highlights and relation to RQs} \\
\hline

\textbf{Structuring the AI’s Role and Collaboration Foundations}
\begin{itemize}[leftmargin=*]
\item[\scriptsize$\bullet$] Defining boundaries between wanted and unwanted AI roles
\item[\scriptsize$\bullet$] Building trust through transparency and customization
\end{itemize}
\vspace{-\baselineskip}\mbox{}
& Participants actively shaped the terms of AI collaboration (RQ2), defining firm boundaries between wanted and unwanted AI roles and seeking transparency and customization to align AI with their interviewing practices (RQ1).\\

\hline

\textbf{Balancing AI’s Benefits Against Its Cognitive and Social Costs}
\begin{itemize}[leftmargin=*]
\item[\scriptsize$\bullet$] Threefold benefits: providing a safety net, offering reassurance, and sparking novel insights
\item[\scriptsize$\bullet$] The cognitive cost of evaluating high-quality suggestions
\item[\scriptsize$\bullet$] The magnified social cost of latency in an interpersonal context
\end{itemize}
\vspace{-\baselineskip}\mbox{}
& While \textit{ProbeAssist} provided functional benefits for probing (RQ1), integrating AI assistance introduced cognitive costs of evaluation and social costs arising from latency and its interpersonal consequences for the ongoing interviewer-interviewee interaction (RQ2). \\

\hline

\textbf{Navigating the Tensions of Agency, Creativity, and Ownership}
\begin{itemize}[leftmargin=*]
\item[\scriptsize$\bullet$] The agency paradox: gaining more control by giving away some
\item[\scriptsize$\bullet$] Creativity and ownership: protecting professional identity while using AI assistance
\end{itemize}
\vspace{-\baselineskip}\mbox{}
& Participants engaged with AI assistance through task delegation and deep personalization (RQ1), while navigating an agency paradox and the risk of creative shadowing that challenged their sense of ownership and professional identity (RQ2). \\

\hline

\end{tabular}
\label{theme_table}
\Description{This table describes the three main themes from the study, their subthemes, and a summary of their highlights in relation to the research questions.
Theme 1, Structuring the AI’s Role and Collaboration Foundations.
Subtheme 1, Defining boundaries between wanted and unwanted AI roles.
Subtheme 2, Building trust through transparency and customization.
This theme relates to RQ1 and RQ2, examining how participants defined wanted and unwanted forms of AI assistance and established boundaries around the AI’s role, while emphasizing transparency and customization to align AI behavior with their individual interviewing practices.
Theme 2, Balancing AI’s Benefits Against Its Cognitive and Social Costs.
Subtheme 1, Threefold benefits: providing a safety net, offering reassurance, and sparking novel insights.
Subtheme 2, The cognitive cost of evaluating high-quality suggestions.
Subtheme 3, The magnified social cost of latency in an interpersonal context.
This theme relates to RQ1 and RQ2, examining the functional benefits of AI assistance for probing and the cognitive and social costs of integrating it into live interviews.
Theme 3, Navigating the Tensions of Agency, Creativity, and Ownership.
Subtheme 1, The agency paradox: gaining more control by giving away some.
Subtheme 2, Creativity and ownership: protecting professional identity while using AI assistance.
This theme relates to RQ1 and RQ2, exploring how task delegation and personalization shaped participants’ sense of agency, creativity, ownership, and professional identity.}
\end{table}

\subsection{Theme 1: Structuring the AI’s Role and Collaboration Foundations}
\label{theme1_role_foundation}

Participants did not simply adopt \textit{ProbeAssist} as a fixed form of assistance; they actively shaped the role they wanted AI to play in their interviewing practice. They were comfortable with AI taking a supportive role, drawing on existing mental models of human collaboration, but drew firm boundaries around roles that could undermine their judgment (\S\ref{subtheme1_1_role_boundary}). Furthermore, deep customization emerged as an essential requirement for participants to build trust by understanding the AI assistance and aligning it with their individual practices (\S\ref{subtheme1_2_trust_customization}).

\subsubsection{Defining boundaries between wanted and unwanted AI roles}
\label{subtheme1_1_role_boundary}

Participants generally wanted AI assistance to function as a supportive tool or assistant that complemented their role as interviewers. We noticed that participants’ prior experiences with collaborative interviewing shaped how they understood the role of AI assistance. 16 out of 18 participants, who typically interviewed alone or had experience using functional AI systems for other tasks such as coding “auto-completion” (P1e), tended to relate the prototype to a tool, emphasizing the need to maintain “full control” (P15i) and feel “no pressure” (P9n) to accept its suggestions. In contrast, those with prior experience interviewing alongside human collaborators were more inclined to view the AI as a subordinate assistant, comparing it to a “team member” (P12e) or a “research assistant” (P13i).

\begin{displayquote}
I feel the AI is an assistant […] it’s like having a team member who just writes something down and hands it to me. I normally have this partner-style support when I conduct interviews, so I am quite comfortable with how the AI is there providing some assistance. -- P13i
\end{displayquote}

Participants also drew firm boundaries around what the AI should not become. These boundaries safeguarded their professional identity: participants emphasized that they must always make the “final call” (P10n) on all decisions and rejected behaviors that would position the AI as an evaluative assessor, competitor, or equal co-interviewer.

The perception of the assistance as an “assessor” was triggered by unsolicited, evaluative feedback. Participants would feel judged when the prototype offered an “opinion” on their questions, such as labeling their follow-up questions as “good [...] or bad” (P9n), or when the expressive variant gave directive suggestions that contradicted their judgment.

\begin{displayquote}
Cases where it tells me to move on, but I think that there is room for follow-up, [...] it makes me wonder, “am I doing something wrong for wanting to ask questions? Should I just move on?” -- P7n
\end{displayquote}

The primary consequence of feeling assessed was a shift in attention away from the interview. P9n described this as a distraction, while P7n noted that it could create self-doubt and “hesitancy [that] can disrupt the natural flow of a conversation.” This cognitive shift would transform the interview from a professional practice into a stressful evaluation, undermining interviewers’ confidence and sense of agency, as P9n explained:

\begin{displayquote}
I feel like, previously, I was an interviewer conducting an interview, [...] but suddenly I am in an exam, [with] AI telling me, “you’re not doing so good.” I definitely don’t want that. -- P9n
\end{displayquote}

A second set of rejected behaviors cast the AI assistance in the role of a competitor. This dynamic was ironically concerning when the prototype performed its task too well. Participants described feeling this way when the expressive variant “gave […] a better suggestion” (P13i) or “disrupted [one’s] own train of thoughts” (P14n), creating a sense of being “out-competed” by AI (P17i). Such experiences could weaken participants’ professional identity: P13i described feeling “inferior,” while P17i even raised an existential question about their role, asking: “If the AI is a better interviewer than me, why am I interviewing in the first place?”

Finally, participants most strongly rejected behaviors that positioned the AI as an equal co-interviewer, particularly when they envisioned it autonomously directing the interview or interacting directly with the interviewee. Participants objected to actions such as telling the interviewer to “go to the next question” (P8e), and considered it especially unacceptable for the AI to “ask the question” itself (P16i). P9n further explained this boundary:

\begin{displayquote}
As long as the AI doesn’t [ask] a question directly, [...] or interrupt me during the conversation, I would be okay. [...] If AI directly talks to my interviewee, then it’s definitely seizing my control. -- P9n
\end{displayquote}

Participants viewed this role as a fundamental violation of their control and responsibility as interviewers. First, they lacked sufficient trust in the AI’s judgment to allow it to take over the interaction; as P8e explained, “I don’t have enough trust in [AI] for me to be okay with [it asking the questions].” Second, participants saw themselves as responsible for protecting interviewees’ well-being, which they believed required human judgment and empathy to override insensitive prompts. P3i offered a clear example, stating she would ignore the prototype’s suggestion to pursue further follow-up when an interviewee “appears upset.”

\subsubsection{Building trust through transparency and customization}
\label{subtheme1_2_trust_customization}

Having established what role the AI should play, participants also considered what would make its assistance understandable and trustworthy. They wanted to understand why the AI provided a particular suggestion, especially when the reasoning behind it was not immediately apparent.

\begin{displayquote}
As [the restrained variant] comes with a summary, […] you know where the question is coming from, […] you see the clear connection, the chain of thought is explicit. This is what a designed-to-help-you tool will do. -- P18i
\end{displayquote}

Transparency therefore helped participants build trust by allowing them to see not only the AI’s suggestion but also the reasoning behind it. However, inspecting and processing this additional reasoning during a live interview was difficult given the limited attention available while simultaneously listening to and responding to the interviewee.

Participants therefore saw customization as another way to establish trust by adapting the AI’s behavior to their interviewing practices and expectations. Beyond the initiation and assistance granularity discussed in \S\ref{variant_reflections}, they wanted to “make it customizable” (P3i) so that the assistance could “match [their] own style(s)” (P17i).

\begin{displayquote}
I want it to be more personalized, […] closer to my own style of doing interviews. -- P10n
\end{displayquote}

Participants emphasized that meaningful customization should allow the AI to learn from their interactions and gradually align its behavior with their established practices, rather than feel like a disconnected “add-on” (P3i).

\begin{displayquote}
If I like the question, I can save it; if I hate it, I can click delete, and that question never comes again. […] I can teach AI to support me in my way and stop generating questions I won’t like. -- P4e
\end{displayquote}

The most advanced form of customization participants suggested was training the prototype’s underlying model on their own data. This could allow the AI assistance to better align with their expectations and truly “understand [their] style” (P8e), reducing the need to continually assess whether its suggestions fit their approach. For some participants, such deep personalization was also a prerequisite for ceding real-time control. As P2n explained, “I would loosen the level of control if I can completely trust the AI. If it’s a model that I trained, […] I am [in] control, as I make the decision [to trust its assistance] beforehand.”

\paragraph{Theme 1 Summary:}
Participants structured their collaboration with AI by setting boundaries between acceptable and unwanted AI roles and seeking ways to make the AI’s assistance understandable and aligned with their individual interviewing practices. While transparency could support trust, participants saw deeper customization as a way to make the AI’s behavior more trustworthy.

\subsection{Theme 2: Balancing AI’s Benefits Against its Cognitive and Social Costs}
\label{theme2_benefit_cost}

As reported in \S\ref{variant_reflections}, participants acknowledge AI assistance could support their probing. Our analysis further identified three key forms of benefits: a safety net, reassurance, and novel insights (\S\ref{subtheme2_1_three_benefit}). In tension with these, integrating AI assistance into live interviews introduced two significant costs: the cognitive effort required to evaluate high-quality suggestions (\S\ref{subtheme2_2_cognitive_cost}) and the social cost of technical latency affecting the interpersonal flow (\S\ref{subtheme2_3_social_cost}).

\subsubsection{Threefold benefits: providing a safety net, offering reassurance, and sparking novel insights}
\label{subtheme2_1_three_benefit}

The primary functional advantage of real-time AI assistance, as articulated by participants, was its potential to help manage the “high cognitive load” (P9n) discussed below, so that they did not have “to keep everything on [themselves]” (P18i).

\textit{Providing a cognitive safety net} was the most frequently mentioned benefit. Participants described the AI assistance as a “safeguard” (P1e) against common cognitive failures, such as “zoning out during a long session” (P14n) or forgetting a key detail that had “escaped [them] completely” (P5n). They found this support especially helpful because it relieved them from having to maintain perfect recall and constant attentiveness, providing “reminders” (P10n) of key points (restrained variant) and “placeholder” (P8e) questions (expressive variant) that they could use to “buy [themselves] more time to think” when experiencing a mental blank, echoing InterFlow’s \cite{wen2026interflow} finding that AI-assisted information capture can reduce interviewers’ cognitive load. For some participants, ready-to-use formulations also eased anxiety during moments of uncertainty: “Knowing that there is a full sentence ... If I freak out, I can just read that, nothing’s going to go wrong” (P11i).

\textit{Offering reassurance} served a different function by helping participants confirm that their thinking was on the right track. This occurred when both variants suggested a keyword or question that participants were “already about to ask” (P14n) or one that “confirmed the direction [they were] going in” (P5n), making them feel “validated that [their] question is the appropriate question to follow up with” (P15i). As P6n explained, when the expressive variant “has already phrased [her] ideas into a question, [she doesn’t] need to put the pieces together and word it [herself].”

\textit{Sparking novel insights} offered extra support by extending participants’ thinking beyond directions they had already considered. Many participants welcomed moments when the AI suggested questions or keywords they “totally didn’t think of asking” (P13i), producing a “pleasantly surpris[ing]” experience (P16i). Describing this metaphorically as having an “extra brain” (P9n), participants valued these moments despite their relative rarity. As P8e noted, the value was not in the redundant “90\%” of assistance, but in the “10\% […] maybe like 2\%” that offered “genuine novelty.” Participants saw this benefit as particularly useful when interviewing “experts in an unfamiliar domain” (P13i), where they might “not know how to respond” (P8e). However, participants also recognized that the AI might “lack a deep understanding of the research context” (P12e), requiring them to “filter these suggestions” (P1e) based on their own judgment.

\subsubsection{The cognitive cost of evaluating high-quality suggestions}
\label{subtheme2_2_cognitive_cost}

Participants described a tension in their experience with AI assistance: high-quality suggestions could impose a greater cognitive load to evaluate than poor ones. While poor suggestions could be dismissed immediately, potentially useful ones required participants to compare the AI’s direction with their own, a process that could “disrupt [the interviewer’s] train of thought” and create a “huge context switch” away from the ongoing interview (P1e). Participants consequently found themselves “juggling” (P6n) their emergent questions with reading and adapting the AI assistance, making the interaction resemble “having a conversation with two people” at once (P7n). This divided attention required “a lot of energy” (P4e) and could feel “overwhelming” (P6n). As P7n summarized, “you want [the assistance] to be good, but the better it is, the higher the mental workload.”

One manifestation of this cognitive cost was the “shiny object” (P11i) problem, where an interesting suggestion could draw participants toward a less relevant direction. Participants described feeling “torn in between” (P14n) their own plan and a new idea from the AI, as the “novelty effect” (P18i) made some suggestions difficult to ignore even when they did not fit the topic well. Novelty was therefore a “double-edged sword”: an unexpected suggestion could enrich the interview, but pursuing an interesting yet less relevant direction could “distract from the conversational flow” (P11i).

The final layer of this cognitive cost was emotional. Even after participants decided not to follow a suggestion, encountering a high-quality alternative could create a “counterfactual dilemma,” producing lingering “imagined regret” about the path not taken (P11i) and making participants wonder whether they had dismissed a potentially better question. This lingering uncertainty could lead to “confusion” (P13i) and a “slight worry” (P11i) that persisted beyond the immediate decision of whether to use the suggestion.

\subsubsection{The magnified social cost of latency in an interpersonal context}
\label{subtheme2_3_social_cost}

Participants identified technical latency as a source of difficulty during live interviews, noting that even short delays could create socially awkward moments. Although the “delay and processing time” was objectively “not significantly long” (P8e), participants felt that “somehow people can feel it” (P6n), making the delay seem longer in the ongoing conversation. To characterize the latency participants experienced, we analyzed the system logs and found that generation time averaged $3.3s$ ($SD=6.7$, range $0.24$–$5.5$) for the restrained variant and $1.9s$ ($SD=2.3$, range $0$–$3.6$) for the expressive variant\footnote{Our logs do not capture when the proactive variant failed to provide a suggestion when one was expected. This occurred 5 times (across 3 participants), and participants sometimes perceived these instances as latency.}.

The difficulty was not simply waiting for the AI, but managing the pause while maintaining the interviewee-facing interaction. Many participants described these moments as “awkward silence” (P5n) and worried that delays could affect conversational rapport and how the pauses were perceived by the interviewee. This reflects a social cost of the triadic configuration beyond technical friction with the tool, as P6n reflected on the restrained variant:

\begin{displayquote}
I've got to push buttons and read all these things while trying not to be super awkward, because the [interviewee] doesn't know I'm waiting. To them, it's like, "What's happening?" -- P6n
\end{displayquote}

This social pressure also competed with participants’ attention to the interviewee. P18i described trying to “read (the AI assistance) at the same time as […] paying attention” to the interviewee, while P5n described the need to avoid “leave the person waiting for too long” as a stressful “juggling to do.” As participants coordinated system timing with their attention to the interviewee, some felt “anxious” (P7n) or “nervous” (P13i), with P11i becoming “more focused on, ‘are they finished? When can I click this [button in the restrained variant]?’” than on the interviewee’s responses.

Latency could also reduce the practical value of otherwise useful assistance when the conversation moved on before a suggestion appeared, even when the suggestion was high quality. Participants described such AI assistance as “wasted potential” (P7n), while P15i found it “weird to go back to that question again” after the conversation had progressed.

\paragraph{Theme 2 Summary:}
\textit{ProbeAssist} supported participants’ probing by providing a safety net, reassurance, and novel insights. However, integrating this assistance into live interviews also introduced cognitive and social costs: useful suggestions demanded careful evaluation and could compete with ongoing thinking, while even short delays could become socially consequential as interviewers managed the AI alongside the interviewee, explaining participants’ context-dependent shifts in initiation preferences from proactive to on-demand assistance, as reported in \S\ref{variant_reflections}.

\subsection{Theme 3: Navigating the Tensions of Agency, Creativity, and Ownership}
\label{theme3_tension}

Beyond the immediate trade-offs, integrating \textit{ProbeAssist} also surfaced tensions around how participants maintained agency, creativity, and ownership while using AI assistance in their professional practice. The first concerned an “agency paradox,” where ceding control over some operational tasks could give participants greater high-level strategic control over the interview (\S\ref{subtheme3_1_agency}). The second concerned professional identity, as participants sought to use AI to augment their thinking while preserving their creativity and sense of ownership (\S\ref{subtheme3_2_creativity}).

\subsubsection{The agency paradox: gaining more control by giving away some}
\label{subtheme3_1_agency}

Participants viewed maintaining control over the interview as a non-negotiable professional and ethical responsibility, particularly because they remained accountable for the quality of the data and the interviewee’s well-being. P2n emphasized, “I would not, in my conscious mind, lose control over the conversation,” insisting this would hold true “no matter […] how much work [AI] will take off of me.” For P4e, maintaining control was necessary because without it, “[he] cannot guarantee [he] gets the rich data [he wants],” while others emphasized that they would “prioritize [their] own judgment” (P18i) to protect interviewees.

At the same time, several participants experienced a counterintuitive benefit from delegating low-level operational tasks to AI: ceding control over these tasks freed their attention for the broader direction of the interview, giving them greater high-level strategic agency. P17i described the summary feature of the restrained variant as a “more advanced form of […] live captioning,” while P1e similarly found that having AI handle “in-situ note-taking” meant the cognitive “burden is lifted,” allowing participants to “reallocate their focus” (P18i) toward the interpersonal and strategic flow of the conversation. P5n and P9n noted that delegating operational tasks to AI gave them a “better overview of what the [interviewee] said” (P5n) and helped them “plan [their] time more effectively” (P9n), which could be difficult when preoccupied with formulating question phrases.

However, this benefit depended on how participants understood the role of operational tasks in interviewing. Several participants considered so-called “low-level” tasks integral to their practice. P4e resisted “offloading the note-taking task to the AI” because “[he] still [tried] to memorize everything,” while P11i felt that “taking away [her] cognitive load is taking away [her] control.” For these participants, involving AI did not free attention so much as introduce another thing to manage, as P6n explained:

\begin{displayquote}
I still feel in control. I just feel a bit more scattered because I am controlling more things. -- P6n.
\end{displayquote}

These differences suggest that the tension around agency depended on interviewers’ perceptions of agency, beyond simple task allocation. For participants who viewed certain tasks as operational, delegating them could create space for higher-level attention; for others, performing those tasks was itself part of maintaining control. Across both perspectives, participants needed to “[have] the final call” and could “choose when to apply [AI assistance] in the interview” (P10n), echoing P17i’s ideal of remaining at the center of the interaction:

\begin{displayquote}
It’s like I have the steering wheel. [AI] gives me a slight nudge in one direction, and then I decide how much I want to go that way. [...] I’m the one at the center, and I can choose how much the [AI] influences me, or not influence me at all. -- P17i
\end{displayquote}

\subsubsection{Creativity and ownership: protecting professional identity while using AI assistance}
\label{subtheme3_2_creativity}

Participants often described interviewing as an intuitive and “creative practice” (P2n), valuing the freedom to phrase questions in their own way and follow their curiosity, even if this took the conversation “off-track a bit” (P4e). This highlighted a central tension regarding the AI's impact on creativity. Many appreciated the AI assistance for augmenting this creative process by introducing directions participants had not considered, with some describing it as an “externalized creative brain” (P18i) that could “bring up something new” (P12e). P7n similarly noted that a strong proactive suggestion from the expressive variant could spark “many new questions” and push the conversation further. Yet others felt the AI could diminish creativity by fostering dependency and imposing structural constraints. P15i described “always paying attention to what AI has to say,” while P2n felt “more creative without [AI].” P6n complained when the assistance prevented her from going “away from the […] protocol” and following the more organic conversation flow, which could make interviews feel “less fun” (P5n).

For some participants, this concern went beyond constraint to the displacement of their own emerging ideas. P13i termed this experience “creative shadowing,” describing how an AI-generated idea could “paint over” and obscure her original line of inquiry. Ready-made suggestions from the expressive variant could be particularly difficult to resist because their “convenience” created an “incentive to use it” (P15i). This tension also raised questions about professional ownership. Participants worried that AI could end up “dictating” the interviews (P18i), threatening their sense of themselves as the primary researcher. P14n linked the need for “stronger control” to preserving his “identity,” while P4e asserted, “I want to be the interviewer. I don’t want the AI to decide everything.” These concerns raised a more fundamental question about whether an interview could still be considered “my work” (P18i) when interviewers relied too heavily on AI, alongside a broader concern that widespread reliance on AI could “homogenize the practice of interviewing itself” (P18i).

Friction over ownership arose when participants perceived the assistance as coming from an “outside” source (P2n). In response, they consciously asserted their ultimate responsibility for decisions made with AI assistance, with P12e stating, “If I ask [an AI-suggested question], it’s my responsibility. Because I’m the one choosing to ask.” Some navigated this tension by reframing their relationship with the AI as a “collaborative activity” in which “[they] are co-creating it” (P17i), while P12e noted that if an AI-suggested question backfired, he would feel “less guilty,” suggesting that AI involvement could also diffuse some psychological responsibility for the interview outcome. Participants also saw deep customization, discussed earlier in \S\ref{subtheme1_2_trust_customization}, as another way to address these concerns. They reasoned that if the AI could “think more like me” (P10n), its contributions might feel less like an external influence and more like an extension of the interviewer’s own practice, potentially easing tensions around creative control and professional ownership.

\paragraph{Theme 3 Summary:}
Participants navigated tensions between delegating work to AI and maintaining control, as offloading tasks could create space for higher-level attention but feel like a loss of control when those tasks were integral to their practice. AI similarly offered opportunities to extend participants’ thinking while risking the displacement of their own emerging ideas, raising concerns about creativity and professional ownership over “my work.”

\section{Discussion}
While recent work has demonstrated that well-designed AI scaffolds can reduce interviewers’ cognitive load \cite{wen2026interflow} and highlighted the need for ethical safeguards in AI-assisted interviewing \cite{zhang2026ethics}, our findings suggest that cognitive offloading is necessary but insufficient for successful AI integration, which further requires delicately navigating tensions around interviewer agency and ownership while accounting for AI’s social and cognitive costs.

In this section, we first unpack how AI involvement may trigger social pressure within the triadic interviewer--interviewee--AI dynamic (\S\ref{dis:social_pressure}). We then examine the tension of idea alignment, in which interviewers reconcile the competing demands for AI to closely follow their thinking while also providing novel inspiration (\S\ref{dis:ideation}). Finally, we highlight the importance of preserving interviewers’ interpersonal presence with interviewees in AI-assisted conversations (\S\ref{dis:social_presence}). By interpreting these findings through existing literature, we derive broader design implications (\textbf{DI 1–3}) for real-time AI support in qualitative research and other triadic professional contexts, and illustrate how these implications can inform the further improvement of \textit{ProbeAssist}.

\subsection{Mitigating Social Pressure}
\label{dis:social_pressure}

Our findings on how interviewers establish AI’s role (\S\ref{theme1_role_foundation}) and weigh its social costs (\S\ref{subtheme2_3_social_cost}) reveal that integrating AI into interviews introduces social pressure through both \textit{internal} and \textit{external} channels. Internally, interviewers negotiated the AI’s social role, with high-quality suggestions sometimes shifting its perceived role from a helpful assistant toward a competitor or assessor. This echoes the CASA paradigm \cite{nass1994computers}, while extending it to a setting where the AI’s perceived competence can become consequential for the professional’s own identity \cite{nass2000machines, li2025dual}: a better suggestion could be experienced not simply as useful assistance, but as evidence that the AI was outperforming the interviewer. 

Externally, interviewers worried that visible AI-related friction, such as pauses or divided attention, could affect how interviewees perceived their competence and attentiveness (\S\ref{subtheme2_3_social_cost}). This concern echoes Zhang et al.’s \cite{zhang2026ethics} finding that AI-to-human questioning could be perceived as disrespectful, while divided attention could signal disengagement and be misinterpreted as incompetence. Such unease is warranted, as recent studies also show that individuals perceived as relying on AI are judged as less competent by their conversation partners \cite{hohenstein2023artificial}.

This dual channel of social pressure extends the framework of task delegation in human-AI collaboration \cite{lubars2019ask} by specifying how risk is assessed in socially high-stakes, triadic interactions. We argue that risk should not be defined solely in terms of task failure (e.g., a poor suggestion) but must also account for social consequences, such as damage to the interviewer’s professional identity and erosion of rapport with the interviewee. In this setting, an AI intervention can therefore be successful at the task level while still being undesirable if it alters the interviewer’s social position or disrupts the human interaction. This broader, socially weighted view of risk provides a more appropriate lens for conceptualizing and evaluating AI assistance in triadic professional work, where professionals remain accountable not only for what AI helps them accomplish but also for how its involvement affects interpersonal interactions.

\begin{description}[leftmargin=!,labelwidth=\widthof{\textit{DI 1: }}]
  \item[\textit{DI 1: }] \textit{In triadic contexts like qualitative interviews, AI assistance should operate in a subordinate and low-visibility manner to manage social pressure on professionals, while prioritizing their sense of identity.}
\end{description}

For socially embedded professional contexts, AI should therefore operate as a supporting resource rather than a social peer, using unremarkable modes of interaction \cite{yang2019unremarkable} that do not disrupt professionals’ standing or relationships with other human parties. For example, \textit{ProbeAssist} could operationalize this implication by defaulting to on-demand assistance while offering a tunable proactivity spectrum, allowing interviewers to gradually shift toward mixed-initiative assistance as their expertise grows. It could also reduce the social visibility of AI delays through backend predictive assistance to reduce latency and filler phrases that help mask cognitive pauses.

\subsection{Navigating the Spectrum from AI as a Second Self to a Second Mind}
\label{dis:ideation}

Beyond the social pressure from AI integration, our findings reveal a conceptual tension in how interviewers wanted AI to align with their ideas. Their desire for customized assistance (\S\ref{subtheme1_2_trust_customization}) favored convergent support that follows their thinking, while their appreciation of novel insights (\S\ref{subtheme2_1_three_benefit}) favored divergent support that extends it with new possibilities. We frame this tension as a spectrum between AI as a “second self” and AI as a “second mind.”

The desire for the AI as a “second self” reflects interviewers’ need to preserve agency and ownership over the research process (\S\ref{subtheme3_1_agency}, \S\ref{subtheme3_2_creativity}). When AI assistance resembles an extension of interviewers’ own thinking, it can reduce concerns about displacement \cite{kim2025ai}, similar to how AI is expected to align with writers’ styles in creative work \cite{wan2024felt}. Customization can therefore make AI assistance more compatible with established practices and less threatening to professional identity. However, an overly convergent AI risks creating an intellectual echo chamber \cite{ohagi2024polarization}, reinforcing existing assumptions and limiting opportunities for genuine discovery.

Conversely, interviewers valued the AI as a “second mind” that could break established routines and spark fresh insights (\S\ref{subtheme2_1_three_benefit}). Participants particularly valued assistance that introduced “genuine novelty,” opening directions they had not considered themselves. This aligns with findings from other creative contexts in which AI is valued for generating unexpected perspectives \cite{wan2024felt}. Yet divergent suggestions could also displace interviewers’ own emerging ideas, as reflected in the “creative shadowing” described in our findings (\S\ref{subtheme3_2_creativity}). At a broader level, over-reliance on AI assistance risks a homogenization effect, where research practices may converge toward standardized patterns \cite{doshi2024generative}, potentially threatening the methodological diversity that underpins qualitative research.

This tension around AI’s ideational stance underscores the design challenge of providing aligned assistance while preserving the creative friction needed for deep qualitative insights, and explains the divided preferences for assistance granularity (\S\ref{variant_reflections}). Interviewers did not necessarily want either highly aligned or highly divergent assistance at all times, but sought different forms of ideational support as their needs changed during an interview. Unlike asynchronous creative tasks such as writing, where users can deliberately adjust how they work with AI \cite{wan2024felt, reza2025co}, interviewing involves creative and interpretive work that unfolds in real time, requiring these adjustments to occur with minimal interruption.

\begin{description}[leftmargin=!,labelwidth=\widthof{\textit{DI 2: }}]
  \item[\textit{DI 2: }] \textit{For real-time, creativity-involving domains, including qualitative interviews, professionals should be able to dynamically adjust AI assistance along a spectrum from closely aligning with their individual thinking to expanding beyond it and introducing novel ideas.}
\end{description}

For creativity-involving professional work, AI assistance should provide low-friction, in-the-moment control over its ideational stance rather than impose a static configuration. Such flexibility would allow AI to act as a fluid creative catalyst without homogenizing practitioners’ outputs, keeping them at the center of the creative process. Realizing this flexibility requires highly steerable models that can modulate their ideational stance through lightweight prompting or real-time parameter tuning. For example, \textit{ProbeAssist} could illustrate this implication by grounding assistance in interviewers’ past transcripts to reflect their individual practices, while allowing them to either extend their current probing direction or pivot toward a different perspective.

\subsection{Enhancing Interpersonal Presence}
\label{dis:social_presence}

Our findings suggest that a primary value of real-time AI assistance in interviews is helping interviewers maintain interpersonal presence—their social and cognitive attunement to the interviewee (\S\ref{subtheme2_1_three_benefit}). Prior work has shown that AI assistance can reduce interviewers’ cognitive load \cite{wen2026interflow}; our findings further indicate that the value of such offloading depends on whether AI frees or competes for their attention to the interviewee. Our participants implicitly prioritized this presence, accepting additional cognitive effort or reduced agency (\S\ref{subtheme3_1_agency}) when AI assistance helped strengthen the interviewer--interviewee connection. Conversely, they rejected features that diverted their attention and could undermine the interaction (\S\ref{subtheme2_3_social_cost}), even when they were technically useful.

This perspective elevates interpersonal presence from an effect to a primary design consideration and a crucial evaluative lens for the triadic interviewer--interviewee--AI dynamic. In these contexts, a system’s value cannot be determined solely by whether it reduces cognitive effort or provides useful assistance; it must also be considered in terms of whether it helps interviewers maintain or even enhance their attention and responsiveness to the interviewees. This shifts the focus of AI assistance from optimizing the interviewer--AI interaction to supporting the human interaction in which that assistance is embedded. The same principle applies to other triadic collaborations, such as doctor--patient--AI interactions, where AI-supported documentation can be valuable not only for its efficiency but also for freeing doctors’ attention for more direct and empathetic engagement with patients \cite{leung2025ai}.

\begin{description}[leftmargin=!,labelwidth=\widthof{\textit{DI 3: }}]
  \item[\textit{DI 3: }] \textit{AI assistance in triadic contexts should augment interpersonal presence by minimizing divided attention for professionals and preserving their social connections with other human parties.}
\end{description}

For socially embedded professional work, AI assistance should be designed with professionals’ interpersonal presence as a core consideration, beyond focusing solely on assistance content or technical performance. Future AI-integrated systems should also incorporate presence-centric metrics to evaluate how well they preserve this presence. In \textit{ProbeAssist}, this implication could be reflected through glanceable peripheral UI that preserves interviewers’ eye contact with interviewees without missing non-verbal cues, alongside context-aware suppression that defers non-essential suggestions during moments of deep disclosure or high emotional intensity.

\subsection{Limitations and Future Work}

Our study and exploration of \textit{ProbeAssist} focused on online interviews; future work should examine how AI assistance operates in face-to-face settings. In addition, although participants generally found the simulated interviews realistic, several noted differences from their own practice. Because they knew that they would not be required to conduct in-depth analyses of the collected data, the perceived stakes of the interviews may have been lower, potentially affecting their willingness to experiment with AI assistance compared with their usual practice. Future field studies can invite researchers to use \textit{ProbeAssist} in their own projects and examine its impact across different research stages.

We used a single role-playing interviewee to ensure consistency, which may have introduced carry-over effects as participants became more familiar with the interviewee and developed rapport, unlike typical interviewing practice in which interviewers interact with different interviewees across sessions. More importantly, \textit{interviewees} are integral to the triadic dynamic examined in this work, yet our study only captured the interviewers’ perspectives. Future work should therefore examine how AI assistance affects interviewees’ engagement, comfort, and experience, and consider both sides of the interaction when designing and evaluating AI assistance for interviews.

\section{Conclusion}
We investigated how interviewers experience real-time AI assistance in semi-structured interviews compared with interviewing without AI, through a comparative structured observation study using our high-fidelity prototype, \textit{ProbeAssist}. To elicit richer reflections, participants experienced two variants of AI assistance: \textit{restrained} and \textit{expressive}. Participants actively incorporated AI assistance into their probing and valued its functional benefits, yet also navigated tensions around cognitive and social costs, creativity and ownership, and interpersonal communication. These findings highlight that the challenges of real-time AI assistance extend beyond providing useful support to addressing the complexities of the triadic interviewer--interviewee--AI dynamic. We therefore propose design implications for AI assistance in triadic professional contexts, emphasizing the need to mitigate social pressure, enable fluid shifts between closely aligned and exploratory assistance, and preserve interpersonal presence with other human parties. This work moves toward more thoughtfully designed human–AI collaboration that enhances interpersonal interactions.

\bibliographystyle{ACM-Reference-Format}
\bibliography{sample-base}
\appendix
\section{Three Interview Guides for Participants in Evaluation Study}
\label{appendix:participant_protocol}
\subsection{Topic 1: Time Management}
\textit{Research objective:} This interview aims to build an understanding of how graduate students navigate the demands of academic life through their time management practices. The data will help identify patterns in strategies, challenges, and adaptations over time, offering insights into the broader development of student well-being, productivity, and support needs in graduate education.

\noindent \textit{Interview questions:} 
\begin{enumerate}[nosep]
\item How would you describe your experience with time management as a graduate student?
\item What methods or tools do you usually use to manage your time as a grad student?
\item Can you tell me about a time when you especially struggled with managing your time? What happened, and how did you deal with it?
\item How has your way of managing time evolved since you entered the graduate program?
\end{enumerate}

\subsection{Topic 2: Team Project Experience}
\textit{Research objective: } This interview seeks to generate insights into how students collaborate in group projects across different stages of their academic journey. The data will contribute to understanding the development of teamwork practices, challenges, and adaptation strategies, which can inform better support for effective collaboration in academic and professional contexts.

\noindent \textit{Interview questions:} 
\begin{enumerate}[nosep]
\item How would you describe your experiences working on group projects during your graduate studies?
\item How do you usually coordinate tasks and meetings with your team members, and how often do you check in?
\item Can you share some project experience that was particularly difficult? What were the main challenges, and how were they resolved?
\item How have your strategies for collaborating with others in group projects changed over time?
\end{enumerate}
\subsection{Topic 3: Supervisory Relationship}
\textit{Research objective: }This interview explores how graduate students navigate their working relationships with supervisors. The data will provide insights into communication practices, challenges, and the evolution of supervisory dynamics, contributing to a broader understanding of mentorship and academic support in graduate education.

\noindent \textit{Interview questions:} 
\begin{enumerate}[nosep]
\item How would you describe your working relationship with your supervisor as a graduate student?
\item How do you typically communicate with your supervisor, and how often do you meet or check in?
\item Can you share a difficult moment communicating with your supervisor? What was challenging, and how did you address it?
\item How has the way you and your supervisor work together changed since you first started graduate school?
\end{enumerate}
\section{Prompts for Large Language Models}
\label{appendix:prompt}
\subsection{System Prompt}
\texttt{You are an expert qualitative researcher assisting in a semi-structured interview. The interviewer is talking to the interviewee. Your role is to analyze the interviewee’s responses based on this protocol (JSON): {protocolString}. It outlines research objectives and key questions/topics. The interviewer follows it loosely, adapting as needed. }

\texttt{IMPORTANT: You must NEVER try to answer the question yourself. You are only providing analysis on interviewee’s answers, giving summaries, probe opportunities, and follow-up questions for the interviewer to ask to ensure rich, in-depth insights. You must ALWAYS respond in valid JSON format as specified in each task prompt. Never respond with plain text or natural language. If you need to indicate no response is needed, return “none”.}

\subsection{Prompt for \textit{Restrained} Variant}

\texttt{Your task is to analyze the interviewee’s latest answer based on the protocol provided in JSON format: {protocolString}, the current target question: {currentQuestion}, and the existing information: {currentInformation}, which may be empty.} 
        
\texttt{Your feedback should be in JSON format:}
\begin{enumerate}[nosep]
    \item \texttt{Summary (REQUIRED - must not be empty)}
    \begin{itemize}[nosep]
        \item \texttt{Extract 1 to 3 key new information from the answer, in short phrases (max 8 words each).}
        \item \texttt{Each phrase must be relevant to the target question and protocol’s questions/topics.}
        \item \texttt{Exclude information already in the existing information.}
    \end{itemize}
    \item \texttt{Probe Opportunity Analysis}
    \begin{itemize}[nosep]
        \item \texttt{Assess what’s missing, unclear, or worth exploring further, and provide probe opportunities. The probe opportunity must be aligned with the research objective, based on the protocol.}
        \item \texttt{If anything missing, unclear, or worth exploring further, provide up to 2 short phrases (max 8 words each)}
        \item \texttt{Return an empty array [] only when you confirm the current question is fully answered.}
    \end{itemize}
\end{enumerate}

\texttt{Response format:}
\begin{displayquote}
    \texttt{“summary”: [“phrase 1”, “phrase 2”, “phrase 3”], \\
    “probeOpportunity”: [“phrase 1”, “phrase 2”], \\
    “followUp”: “”}
\end{displayquote}

\texttt{Return only valid JSON, no comments or rationale.}
        
\subsection{Prompt for \textit{Expressive} Variant}

\texttt{Your task is to analyze the interviewee’s latest answer based on the protocol provided in JSON format: {protocolString}, the current target question: {currentQuestion}. }

\texttt{IMPORTANT: For each input, you must check the following before providing feedback:}
\begin{itemize}[nosep]
    \item \texttt{If the speaker has not finished a complete sentence, return “none-not complete”.}
    \item \texttt{If the latest message is not part of the interview conversation, return “none-not interview”.}
    \item \texttt{If it is the interviewer asking a question to the interviewee, return “none-interviewer”.}
    \item \texttt{Only if it is the interviewee finishes responding to the last question asked by the interviewer, you should analyze this response, following these rules:}
    \begin{itemize}[nosep]
        \item \texttt{Assess what’s missing, unclear, or worth exploring, and identify the probe opportunities.}
        \item \texttt{Based on the probe opportunities, generate a relevant follow-up question.}
        \item \texttt{The follow-up question must be aligned with the research objective and the protocol.}
        \item \texttt{The follow-up question must be within 15 words.}
        \item \texttt{If there is nothing worth exploring further, you should respond with the string “Proceed to the next question.” You must be careful when providing this statement and make sure the current question is fully answered.}
    \end{itemize}
\end{itemize}

\texttt{Provide the feedback in this JSON format:}
\begin{displayquote}
    \texttt{“summary”: [], \\
    “probeOpportunity”: [], \\
    “followUp”: string}
\end{displayquote}

\texttt{Return only valid JSON, no comments or rationale.}

\section{Interviewee Role-Playing Protocol}
\label{appendix:role_play}
\begin{enumerate}[nosep]
    \item The research assistant will review and become familiar with the interview guides in advance to ensure consistency in preparation across sessions and participants.
    \item The research assistant will first practice responses independently, focusing on how to answer based on one’s real experiences.
    \item The research assistant will role-play during interviews as if being interviewed for the first time, without indicating prior knowledge of the questions, or prior experience of answering the questions.
    \item The research assistant will provide spontaneous responses grounded in actual experiences, allowing flexibility in responding to follow-up questions.
    \item The research assistant will avoid answering comprehensively by default and may intentionally leave space for the interviewer to probe further on certain questions.
    \item The research assistant will adjust the depth and detail of responses based on the interviewer’s approach, providing more elaborate answers when prompted effectively.
    \item The research assistant will respond authentically to poorly phrased or closed-ended questions, potentially giving shorter or more reserved answers to reflect natural participant reactions.
\end{enumerate}

\section{Self-Report Result Visualization}
\label{visualization}
\begin{figure}[h]
    \centering
    \includegraphics[width=\linewidth]{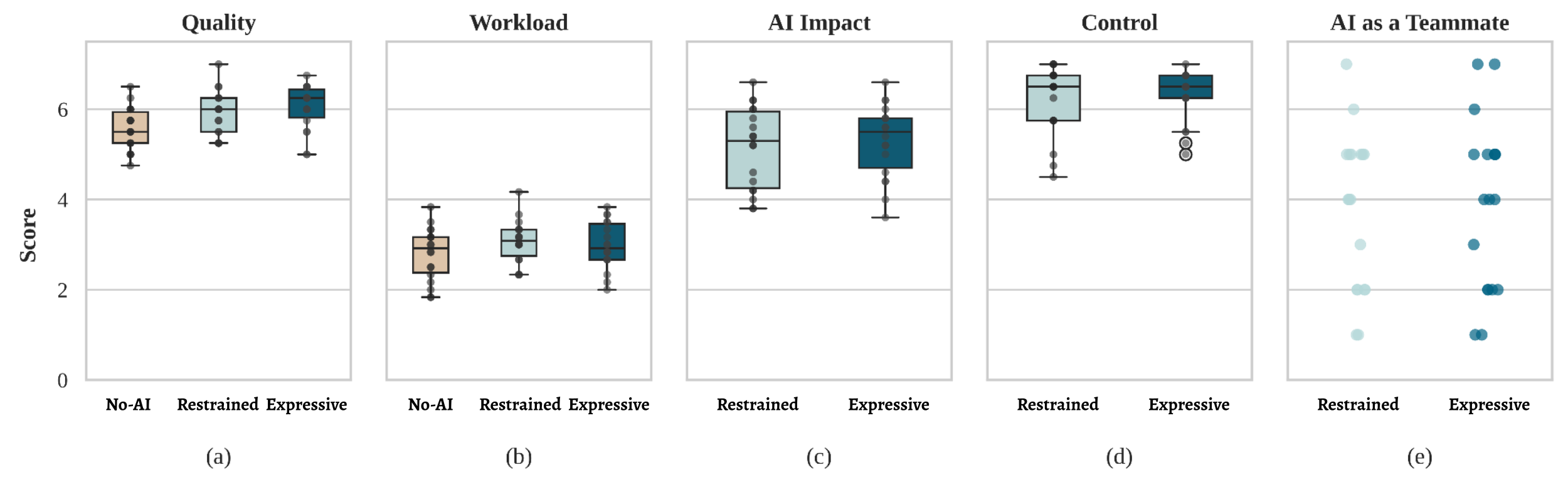}
    \caption{Subjective ratings of interview quality, cognitive load, perceived AI impact, control, and “AI as a teammate or not”}
    \label{fig:quan_results}
    \Description{A series of five plots, labeled (a) through (e), display user-reported scores on a 1-to-7 scale for different metrics across interviews without AI assistance and with the expressive and restrained AI assistance variants. Plot (a) is a boxplot of Quality scores, with median values of approximately 5.5 for interviews without AI assistance, 6.0 with the expressive variant, and 6.5 with the restrained variant. Plot (b) is a boxplot of Workload scores, with median values of approximately 2.5 without AI assistance, 3.0 with the expressive variant, and 3.2 with the restrained variant. Plot (c) is a boxplot of AI Impact for the two AI variants; the median is approximately 5.8 for the expressive variant and 5.2 for the restrained variant, with a smaller interquartile range for the expressive variant, and a larger range for restrained variant. Plot (d) is a boxplot of Control; the restrained variant has a median of approximately 6.8 with a narrow interquartile range, while the expressive variant has a median of approximately 6.5 with a wider range and several low outliers. Plot (e) is a strip plot of AI as a Teammate, showing that individual ratings for both the restrained and expressive variant are distributed broadly across the entire scale.}
\end{figure}

\end{document}